\documentclass[12pt,a4paper]{article}
\pdfoutput=1
\usepackage{color}
\usepackage{amssymb,amsmath,bbold,MnSymbol}
\usepackage{epsf}
\usepackage{epsfig}
\usepackage{relsize}
\usepackage{mathtools}
\usepackage[dvipsnames]{xcolor}
\usepackage[linktoc=page,bookmarks=false,colorlinks=false,linkbordercolor=RoyalBlue,citebordercolor=ForestGreen,urlbordercolor=CornflowerBlue]{hyperref}
\usepackage{latexsym,mathrsfs,dsfont}
\usepackage[normalem]{ulem} 
\usepackage[compress]{cite}
\usepackage{graphicx}
\usepackage{url}
\usepackage{booktabs}
\usepackage{float}
\usepackage{multirow}
\usepackage{changepage}
\usepackage[hypcap]{caption, subcaption}

\usepackage[titles]{tocloft}

\usepackage{tabularx,colortbl}

\usepackage{fancyhdr}
\advance \headheight by 3.0truept       

\allowdisplaybreaks[1]

\definecolor{red}{cmyk}{0,1,1,0.4}
\definecolor{darkgreen}{rgb}{0.0,0.6,0.0}
\definecolor{cDarkGrey}{RGB}{91,91,91}
\definecolor{cGrey}{RGB}{245,243,238}
\definecolor{cBlue}{RGB}{0,110,191}
\definecolor{cLightBlue}{RGB}{214,237,252}
\definecolor{cRed}{RGB}{196,0,100}
\definecolor{cLightRed}{RGB}{254,222,237}
\definecolor{cGreen}{RGB}{0,166,80}
\definecolor{cLightGreen}{RGB}{254,222,237}
\definecolor{cOrange}{RGB}{221,74,44}
\definecolor{cLightOrange}{RGB}{255,215,210}
\definecolor{cPurple}{RGB}{93,35,125}
\definecolor{cLightPurple}{RGB}{241,230,252}
\definecolor{cYellow}{RGB}{252,191,10}
\definecolor{cISSRBlue}{RGB}{0,111,174}
\definecolor{cISSRGrey}{RGB}{167,169,172}

\newcommand{\beq}{\begin{equation}}
\newcommand{\eeq}{\end{equation}}
\newcommand{\be}{\begin{equation}}
\newcommand{\ee}{\end{equation}}
\newcommand{\bi}{\begin{itemize}}
\newcommand{\ei}{\end{itemize}}
\newcommand{\ba}{\begin{array}}
\newcommand{\ea}{\end{array}}
\newcommand{\beqa}{\begin{eqnarray}}
\newcommand{\eeqa}{\end{eqnarray}}
\newcommand{\bea}{\begin{eqnarray}}
\newcommand{\eea}{\end{eqnarray}}
\newcommand{\beqn}{\begin{eqnarray}}
\newcommand{\eeqn}{\end{eqnarray}}

\newcommand{\nn}{\nonumber}

\newcounter{TODO}

\newcommand{\oL}[1]{\overline{#1}}

\newcommand{\ord}[1]{\mathcal{O}\left( #1 \right)}

\newcommand{\DF}{\Delta F}

\newcommand{\bm}{\boldmath}

\newcommand{\alS}{\alpha_s}

\newcommand{\EV}[2][{}]{E_{#2}^{\text{#1}}}
\newcommand{\OpL}[2][{}]{Q_{#2}^{#1}}
\newcommand{\opL}[3][{}]{[O_{#2}^{#1}]_{#3}}

\newcommand{\OpLt}[2][{}]{\widetilde{Q}_{#2}^{#1}}

\newcommand{\Nf}{N_f}

\begin{document}

\begin{flushleft}
{\em \today}
\end{flushleft}

\vspace{-14mm}
\begin{flushright}
 {AJB-22-1}
\end{flushright}

\medskip

\begin{center}
{\Large \bf\boldmath
  Simple Rules for Evanescent Operators in One-Loop Basis Transformations
}
\\[1.2cm]
{\bf
  Jason~Aebischer$^{a}$,
  Andrzej~J.~Buras$^{b}$ and
  Jacky Kumar$^{b}$
}\\[0.5cm]

{\small
$^a$Physik-Institut, Universit\"at Z\"urich, CH-8057 Z\"urich, Switzerland \\[0.2cm]
$^b$TUM Institute for Advanced Study,
    Lichtenbergstr. 2a, D-85747 Garching, Germany \\[0.2cm]
}
\end{center}

\vskip 1.0cm

\begin{abstract}
\noindent

Basis transformations often involve Fierz and other relations which are only valid in $D=4$
dimensions. In general $D$ space-time dimensions however, evanescent operators have to be introduced, in order to preserve such identities. Such evanescent operators contribute to one-loop basis transformations as well as to two-loop renormalization group running. We present a simple procedure on how to systematically change basis at the one-loop level by obtaining shifts due to evanescent operators. 
As an
example we apply this method to derive the one-loop basis transformation from the 
BMU (Buras, Misiak and Urban) basis useful for NLO QCD calculations, to the JMS (Jenkins, Manohar and Stoffer)  basis used in the matching to the SMEFT.

\end{abstract}

\setcounter{page}{0}
\thispagestyle{empty}
\newpage

\setcounter{tocdepth}{2}
\setlength{\cftbeforesecskip}{0.21cm}
\tableofcontents

\newpage

%
%
%
\section{Introduction}

\label{sec:intro}

In recent years a lot of progress has been made concerning Next-to-Leading Order (NLO) analyses, which involve one-loop matching calculations as well as solving two-loop Renormalization Group Equations (RGEs) for the Wilson coefficients of Effective Field Theories. For an up to date review see \cite{Buras:2020xsm}. For instance, concerning the Standard Model Effective Theory (SMEFT), the full matching from the SMEFT onto the Weak Effective Theory (WET) valid below the electroweak (EW) scale is known at tree-level \cite{Jenkins:2017jig} and since recently also at the one-loop level \cite{Dekens:2019ept,Aebischer:2015fzz}. Furthermore, the one-loop RGEs in the SMEFT \cite{Jenkins:2013zja,Jenkins:2013wua,Alonso:2013hga} and in the WET \cite{Jenkins:2017dyc,Aebischer:2017gaw} are known.

In the process of performing a NLO analysis, it is often necessary to perform one-loop transformations between different operator bases, since for instance anomalous dimension matrices (ADMs) are known
only in a particular basis, whereas the matching conditions are given in {a} different one. This is for example the case for the WET, where the two-loop ADMs are known in the BMU (Buras, Misiak and Urban) basis \cite{Buras:2000if} as elaborated on recently in \cite{Aebischer:2021raf}. On the other hand the one-loop matching from SMEFT onto WET is given in the JMS (Jenkins, Manohar and Stoffer) basis defined in \cite{Jenkins:2017jig}.\footnote{We follow here the \texttt{WCxf} convention defined in \cite{Aebischer:2017ugx}.}

Generally, in order to translate the results from one basis to another one at NLO, one-loop basis transformations have to be taken into account. In this respect particular care has to be taken, if the tree-level (Leading Order (LO)) transformations involve Dirac space identities, which are only valid in four space-time dimensions. Examples include Fierz identities or identities involving gamma matrices. When using dimensional regularization, where the divergent loop integrals are continued to $D$ dimensions, such identities can not be used directly, but need to be generalized by introducing evanescent (EV) operators \cite{Buras:2000if}. Such evanescent operators vanish in four dimensions to conserve the original identities but are non-zero in $D$ dimensions. They are therefore formally speaking proportional to $\epsilon = (4-D)/2$, which implies that they give non-zero contributions when inserted into divergent loop diagrams. These are exactly the contributions which enter basis transformations at the one-loop order. In this article we discuss a simple procedure on how to obtain these contributions by computing one-loop corrections resulting from the presence of the evanescent operators.\footnote{A more formal procedure on how to perform NLO basis changes can be found in \cite{Buras:2000if}.} To this end the evanescent operators are simply defined as the difference between operators and their transformed versions using $D=4$ identities. The resulting contributions will manifest themselves in shifts in the corresponding Wilson coefficients of the initial operators.

Having a simple algorithm at hand to perform NLO basis changes is important when performing one-loop matching calculations or in the computation of two-loop ADMs. In this work we explain the underlying formal framework and provide such an algorithm, based on Greek projections \cite{Tracas:1982gp}, which facilitates the aforementioned calculations. In this manner this procedure is an important ingredient in the pursuit of a complete NLO SMEFT analysis.

The rest of the article is organized as follows: In Section~\ref{sec:procedure} we outline the general procedure on how to compute one-loop basis transformations between two operator bases. In Section~\ref{sec:JMStoBMU} we show an explicit example by performing a NLO change from the BMU to the JMS basis.
 {In Section~\ref{sec:calc} we define and calculate the EV operators which in turn gives us the transformation matrix between BMU and JMS bases at the one-loop level.} 
Finally we conclude in Section~\ref{sec:conclusion}. Additional material used in the calculations is collected in the appendices.

\section{Procedure}\label{sec:procedure}
In this section we discuss the full NLO basis transformation between general operator bases. However, in order to make the subsequent sections more
  transparent we will dub the two operator bases JMS (see Table.~\ref{tab:jms} for the list of operators considered 
in this work) and BMU.
We start with the simple LO basis transformation which will also set the notation used. Then, in the second subsection we discuss the issue of evanescent operators which become relevant when performing  basis changes at the one-loop level.

\begin{table}[t]
\centering
\renewcommand{\arraystretch}{1.5}
\begin{tabular}{||c|c||c|c||}
\hline\hline
\multicolumn{2}{||c||}{$(\bar LL)(\bar LL)$} & \multicolumn{2}{|c||}{$(\bar RR)(\bar RR)$}  \\ \hline
$\opL[V,LL]{dd}{prst}$ & $(\bar{d}_L^p \gamma_\mu d_L^r) (\bar{d}_L^s \gamma^\mu d_L^t)$ &
$\opL[V,RR]{dd}{prst}$& $(\bar{d}_R^p \gamma_\mu d_R^r) (\bar{d}_R^s \gamma^\mu d_R^t)$  \\
$\opL[V1,LL]{ud}{prst}$ & $(\bar{u}_L^p \gamma_\mu u_L^r) (\bar{d}_L^s \gamma^\mu d_L^t)$ &
$\opL[V1,RR]{ud}{prst}$ & $(\bar{u}_R^p \gamma_\mu u_R^r) (\bar{d}_R^s \gamma^\mu d_R^t)$  \\
$\opL[V8,LL]{ud}{prst}$& $(\bar{u}_L^p \gamma_\mu T^A u_L^r) (\bar{d}_L^s \gamma^\mu T^A d_L^t)$  &
$\opL[V8,RR]{ud}{prst}$& $(\bar{u}_R^p \gamma_\mu T^A u_R^r) (\bar{d}_R^s \gamma^\mu T^A d_R^t)$ \\  \hline\hline
\multicolumn{2}{||c||}{$(\bar LL)(\bar RR)$} & \multicolumn{2}{|c||}{$(\bar LR)(\bar LR)$+ \text{h.c.}}  \\ \hline
$\opL[V1,LR]{dd}{prst}$&$(\bar{d}_L^p \gamma_\mu d_L^r) (\bar{d}_R^s \gamma^\mu d_R^t)$&
$\opL[S1,RR]{dd}{prst}$& $(\bar{d}_L^p d_R^r) (\bar{d}_L^s d_R^t)$\\
$\opL[V8,LR]{dd}{prst} $&$(\bar{d}_L^p \gamma_\mu T^A d_L^r) (\bar{d}_R^s \gamma^\mu T^A d_R^t)$&
$\opL[S8,RR]{dd}{prst}$& $(\bar{d}_L^p T^A d_R^r) (\bar{d}_L^s T^A d_R^t)$\\
$\opL[V1,LR]{ud}{prst}$&$ (\bar{u}_L^p \gamma_\mu u_L^r) (\bar{d}_R^s \gamma^\mu d_R^t)$&
$\opL[S1,RR]{ud}{prst}$& $(\bar{u}_L^p u_R^r) (\bar{d}_L^s d_R^t)$\\
$\opL[V8,LR]{ud}{prst}$&$(\bar{u}_L^p \gamma_\mu T^A u_L^r) (\bar{d}_R^s \gamma^\mu T^A d_R^t) $&
$\opL[S8,RR]{ud}{prst}$& $(\bar{u}_L^p T^A u_R^r) (\bar{d}_L^s T^A d_R^t)$\\
$\opL[V1,LR]{du}{prst}$&$(\bar{d}_L^p \gamma_\mu d_L^r) (\bar{u}_R^s \gamma^\mu u_R^t)$&
$\opL[S1,RR]{uddu}{prst}$& $(\bar{u}_L^p d_R^r) (\bar{d}_L^s u_R^t)$\\
$\opL[V8,LR]{du}{prst}$&$(\bar{d}_L^p \gamma_\mu T^A d_L^r) (\bar{u}_R^s \gamma^\mu T^A u_R^t)$&
$\opL[S8,RR]{uddu}{prst}$& $(\bar{u}_L^p T^A d_R^r) (\bar{d}_L^s T^A u_R^t)$\\
$\opL[V1,LR]{uddu}{prst}$&$(\bar{u}_L^p \gamma_\mu d_L^r) (\bar{d}_R^s \gamma^\mu u_R^t)$ + h.c.&
& \\
$\opL[V8,LR]{uddu}{prst}$&$(\bar{u}_L^p \gamma_\mu T^A d_L^r) (\bar{d}_R^s \gamma^\mu T^A u_R^t)$ + h.c.&
& \\
\hline \hline
\end{tabular} 
\caption{\small Non-leptonic $\DF=1$ operators (baryon and lepton number conserving) in the JMS basis \cite{Jenkins:2017jig}. Note that $\OpL[V1,LR]{uddu}$ and $\OpL[V8,LR]{uddu}$ have Hermitian conjugates. The same holds for the
operators $(\oL{L}R)(\oL{L}R)$. This choice of basis eliminates all
operators with Dirac structures $\sigma^{\mu\nu}$. The class of operators
$(\oL{L}R)(\oL{R}L) + \text{h.c.}$ does not contain non-leptonic operators,
but only semi-leptonic ones.
}
\label{tab:jms}
\end{table}

\subsection{Tree-level transformation}
Let us consider the two bases
\be
\vec{\mathcal{O}}_{\rm BMU}=\{Q_1,Q_2,...,Q_N\}, \qquad \vec{\mathcal{O}}_{\rm JMS}=\{O_1,O_2,...,O_N\},
\ee
containing $N$ operators each. At tree-level each operator is given by a linear combination of operators from the other basis:
\be\label{eq:tree}
\vec{\mathcal{O}}_{\rm JMS}  = \hat R^{(0)}\,  \vec{\mathcal{O}}_{\rm BMU},
\ee
where the $N\times N$ matrix $\hat R^{(0)}$ denotes the linear transformation between the two bases. The superindex '$(0)$' denotes the tree-level transformation in anticipation of the one-loop transformation discussed in the next subsection. The matrix $\hat R^{(0)}$ is obtained by applying identities such as Fierz relations and gamma matrix identities to the operators on the LHS of eq.~\eqref{eq:tree}. It is independent of the renormalization scale and only contains numerical factors.\footnote{This is true up to possible normalization factors, which can contain coupling constants or other parameters that depend on the renormalization scale.}

As seen in eq.~(\ref{eq:tree}) the BMU operator basis is transformed via $\hat R^{(0)}$ into the JMS basis.
This transformation is useful because the hadronic  matrix elements of operators
are usually calculated in the BMU basis and this transformation allows to obtain
them in the JMS basis. However, the Wilson coefficients are nowadays
calculated in the JMS basis, since the SMEFT matching results are only available in that particular basis. Therefore for Wilson coefficients the transformation from JMS to BMU is more useful, which reads
\be\label{eq:treeWC}
\vec{\mathcal{C}}^{(0)}_{\rm BMU}= (\hat R^{(0)})^T\,\vec{\mathcal{C}}^{(0)}_{\rm JMS}\,,
\ee
that is $(\hat R^{(0)})^T$ is involved.

Since $ \hat R^{(0)}$ is invertible by definition, one can
easily express the BMU operators in terms of JMS ones and the JMS Wilson coefficients in terms of BMU ones. 
For our purposes however, the transformations
in eqs.~(\ref{eq:tree}) and (\ref{eq:treeWC}) are more convenient.

{It should be stressed that although the procedure outlined below is general and can be used for any basis transformation the explicit values of various
  coefficients  are given in our paper in the NDR-$\overline{\text{MS}}$ scheme as defined in \cite{Buras:1989xd} with
evanescent operators entering two-loop calculations defined by the so-called Greek method. The details in the
context of WET and SMEFT are discussed in Appendix E of \cite{Aebischer:2020dsw}.}

\subsection{One-loop transformation}
At the one-loop level special care has to be taken when identities have been used in the LO transformation, that are only valid in $D=4$ space-time dimensions. Therefore, for all the lines in $R^{(0)}$ which were obtained using $D=4$ relations, evanescent operators have to be introduced to generalize these identities. In the case of the BMU$\rightarrow$JMS transformation we proceed as follows:

{\bf Step 1:}

We perform a Fierz transformation on  every operator in the BMU basis $Q_i$
and denote the result of this transformation by $\widetilde Q_i$ given
generally by
\be\label{Qtilde}
{\widetilde Q_i = \sum_k \omega_{ik}Q_k\,,}
\ee
with operators $Q_k$ belonging to the BMU basis and coefficients $\omega_{ik}$ determined
through the Fierz identities collected in the Appendix~\ref{subsec:fierz}.
{It should be emphasized that the relation above and analogous relations below should  be interpreted as effective contributions from this operator in one
  loop diagrams. This means that the calculations performed in two bases, in our case BMU and JMS, will give the same results for physical observables when this transformation is taken into account
  at all stages, therefore when calculating also anomalous dimensions. This relations
  and analogous relations involving evanescent  operators can also be used
  for the calculations of two-loop anomalous dimensions of operators
  because there the evanescent operators contributing to two-loop
  anomalous dimensions appear in one-loop sub-diagrams \cite{Buras:1989xd}.
  In order to go to the next order in perturbation theory the shifts should
  include corrections of order $\ord{\alpha_s^2}$ that can be obtained through insertions of
  the operators into two-loop diagrams. }

{\bf Step 2:}

We insert $Q_i$ and $\widetilde Q_i$ defined by (\ref{Qtilde}) into current-current and QCD penguin diagrams of Figs.~\ref{fig:cc-ins} and \ref{fig:peng-ins}, respectively. Due to the presence of evanescent operators which are simply
defined by
\be\label{eq:EV}
  Q_i = \widetilde Q_i +\text{EV}_i\,,
\ee
the insertion of $Q_i$ and $\widetilde Q_i$ into one-loop diagrams will generally differ at $\mathcal{O}(\alpha_s)$, leading to the result
\be\label{FR}
  Q_i = \widetilde Q_i +\frac{\alpha_s}{4\pi} \sum_r {\tilde\omega_{ir}} Q_r,
\ee
where the operators in the sum can again be written in the BMU basis but are generally
different from the ones in the definition of $\widetilde Q_i$ in (\ref{Qtilde})
and also {$\omega_{ik}\not=\tilde\omega_{ir}$.}
Note that this  can easily be done at the one-loop level because in transforming the operators
entering these corrections $D=4$ Fierz identities can be used as will be explained below.
Expressing the shift in terms of BMU operators allows to use
these results for transformations of the BMU basis to any basis, not just the JMS one.

To compute the $\alpha_s$-corrections resulting from the evanescent operator $\text{EV}_i$ in \eqref{eq:EV}, one simply inserts the difference $Q_i - \widetilde Q_i$ into the relevant one-loop diagrams. Since the evanescent operator is formally $\mathcal{O}(\epsilon)$, only the divergent pieces of the loop integrals will contribute and consequently the finite pieces can be discarded in the calculation. 

{The relation between operators in JMS basis and the Fierz transformed operators in BMU are further discussed in Sec.\ref{sec:treeloop}. In particular the tree-level relations are given in eqs.~\eqref{eq:rmatrix}-\eqref{eq:R0VLR}, and the one-loop relations in eqs.~\eqref{eq:R1}-\eqref{eq:R1SRR}.}

{\bf Step 3:}

Having the relations in (\ref{FR}) and inspecting which Fierz
transformations had to be performed to find the LO matrix $\hat R^{(0)}$
we can generalize  the basis change matrix $\hat R^{(0)}$ to the one-loop order, as follows:\footnote{In the case JMS$\leftrightarrow$BMU we will focus on QCD corrections. The relation in eq.~\eqref{eq:basic} can easily be generalized to include other one-loop corrections.}
\begin{equation}\label{eq:basic}
  \vec{\mathcal{O}}_{\rm JMS}  = \hat R\,  \vec{\mathcal{O}}_{\rm BMU}\,,\qquad
  \hat R = \hat R^{(0)}+ \frac{\alpha_s}{4\pi}\hat R^{(1)}\,,
\end{equation}
where $\hat R^{(1)}$ is the one-loop basis transformation matrix resulting exclusively from the evanescent operators.

When computing the one-loop corrections to the difference $Q_i - \widetilde Q_i$, new Dirac structures can appear, which are not present in the original basis, in
our case the BMU basis.
To reduce these structures to the ones in $\vec{\mathcal{O}}_{\rm BMU}$ it is essential
  to use the same projections as in the calculations of two-loop anomalous
  dimensions of the operators. Only then are the evanescent operators
  entering the two-loop calculations the same as the ones used in the one-loop
  matching between WET and SMEFT and in particular they correspond to the ones in the basis change. This prescription guarantees that the renormalization scheme dependence of two-loop
  matrix elements can be canceled by the one present in one-loop contributions
  so that the physical amplitudes are renormalization scheme independent.

  These issues have been discussed in the context of the  NLO QCD calculations  of Wilson coefficients at length in \cite{Buras:1989xd,Dugan:1990df,Herrlich:1994kh} and the summary can be found in Section 5.2.9 of \cite{Buras:2020xsm}.
  There the so-called Greek projection \cite{Tracas:1982gp}, properly
  generalized to include evanescent operators in \cite{Buras:1989xd}, has
  been discussed in detail. As pointed out in \cite{Herrlich:1994kh} this is not
  the only way to include evanescent operators but in fact the simplest one.
  It has been used in all two-loop calculations performed by the second
  author and will be used in the following but this time in the context
  of basis change. It should be kept in mind that this procedure defines the evanescent operators
     in the NDR-$\overline{\text{MS}}$ scheme combined with  the Greek projections as used 
in \cite{Buras:1989xd}. While giving the same results it  is  much simpler than the formal method
  presented in \cite{Buras:2000if} and used recently in \cite{Aebischer:2021raf}.

The NLO transformation in terms of the Wilson coefficients is given as  follows.
 \be
      \vec{\mathcal{C}}_{\rm BMU}=\hat R^T \vec{\mathcal{C}}_{\rm JMS}.
      \ee
Writing
      \be
      \vec{\mathcal{C}}_{\rm JMS}=\vec{\mathcal{C}}^{(0)}_{\rm JMS}+\frac{\alpha_s}{4\pi}\vec{\mathcal{C}}^{(1)}_{\rm JMS}\,, \qquad \vec{\mathcal{C}}_{\rm BMU}=\vec{\mathcal{C}}^{(0)}_{\rm BMU}+\frac{\alpha_s}{4\pi}\vec{\mathcal{C}}^{(1)}_{\rm BMU}\,,
      \ee
the transformation for Wilson coefficients reads
      \be
      \vec{\mathcal{C}}^{(0)}_{\rm BMU}=(\hat R^{(0)})^T\vec{\mathcal{C}}^{(0)}_{\rm JMS}\,,\qquad \vec{\mathcal{C}}^{(1)}_{\rm BMU}=(\hat R^{(0)})^T\vec{\mathcal{C}}^{(1)}_{\rm JMS}+(\hat R^{(1)})^T\vec{\mathcal{C}}^{(0)}_{\rm JMS}\,.
      \ee

{A few remarks concerning the generality of our results are in order: The procedure laid out in this article is not limited to the BMU and JMS basis, but is valid for any pair of operator bases, that are related via Fierz transformations. The full set of relevant evanescent operators is generated by applying Greek identities, or any other identities to reduce the resulting Dirac structures from the one-loop calculation. This choice of Dirac reduction, together with the chosen finite counterterms and the treatment of $\gamma_5$ fixes the renormalization scheme \cite{Herrlich:1994kh}. In this article we have chosen the NDR-$\overline{\text{MS}}$ scheme in combination with the Greek identities. Choosing a different renormalization scheme would change the entries of the rotation matrix $R^{(1)}$ in eq.~\eqref{eq:basic}, but is related via a trivial change of scheme to our results \cite{Chetyrkin:1997gb,Gorbahn:2004my}. Furthermore, the procedure is independent of possible group theory relations between the two bases, since these are still valid in $D\neq 4$ space-time dimensions, and therefore do not obtain any one-loop shifts from evanescent structures. Finally, we note that the operator shifts presented in this paper can be interpreted as one-loop corrections to the original Fierz identities. This has been shown in two recent publications \cite{Aebischer:2022aze,Aebischer:2022rxf}, in which the shifts for all possible four-fermi operators together with the contributions from dipole operators have been taken into account. A similar procedure in the SMEFT has been employed in \cite{Fuentes-Martin:2022vvu}.}

\subsection{How to use this procedure}
\label{subsec:procedure}
Having the results in (\ref{FR}) to be presented in the next section, our goal will be to find the matrix $\hat R$.
To this end comparing the BMU and JMS bases one has to find
those  operators or groups of them for which a Fierz transformation
on operators in the BMU basis has to be performed in order to obtain
the operators in the JMS basis with order $\alpha_s$ corrections taken into account.

Generally the matrix $\hat R^{(0)}$ will have a block structure so that
operators in a given block can be separately considered from other blocks.
  In this context the following three cases arise.
  \begin{itemize}
  \item
    If no Fierz transformations are required in a given block the
    corresponding matrix
    $\hat R^{(1)}$ will vanish and tree-level
    results will be valid also at one-loop. One can use the corresponding
    block in $\hat R^{(0)}$
    in that case.
  \item
    If Fierz transformations in a given block are  required but the contributions of evanescent operators will vanish the corresponding block in the tree-level matrix $\hat R^{(0)}$ will again represent the corresponding block in the full $\hat R$.
    \item Finally, in certain blocks the necessity of performing Fierz transformations will introduce evanescent operators which will contribute to $\hat R^{(1)}$.
      \end{itemize}

  In the next section we will present the three step procedure
  outlined in this section  in explicit terms.

\section{BMU to JMS translation at one-loop}\label{sec:JMStoBMU}
%
\subsection{Basic method}\label{BasicM}
\label{subsec:brules}
\begin{figure}[H]
\centering
  \begin{subfigure}[t]{0.32\textwidth}
    \centering
    \includegraphics[trim={5cm 20cm 12.9cm 4cm},clip,width=0.7\textwidth]{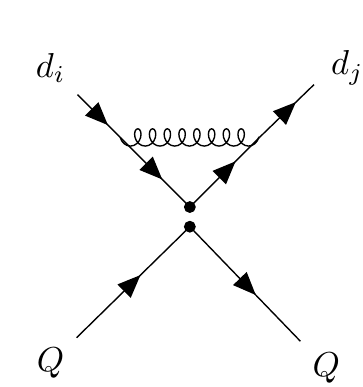}
    \caption{}
    \label{fig:CC-ins-1l}
  \end{subfigure}
  \begin{subfigure}[t]{0.32\textwidth}
    \centering
    \includegraphics[trim={5cm 20cm 12.9cm 4cm},clip,width=0.7\textwidth]{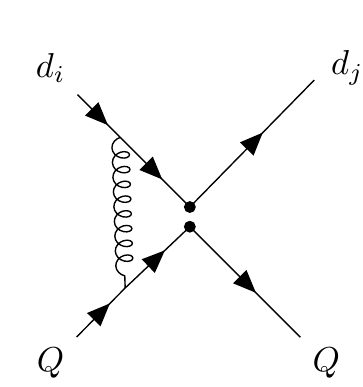}
    \caption{}
    \label{fig:CC-ins-A-1l}
  \end{subfigure}
 \begin{subfigure}[t]{0.32\textwidth}
    \centering
    \includegraphics[trim={5cm 20cm 12.9cm 4cm},clip,width=0.7\textwidth]{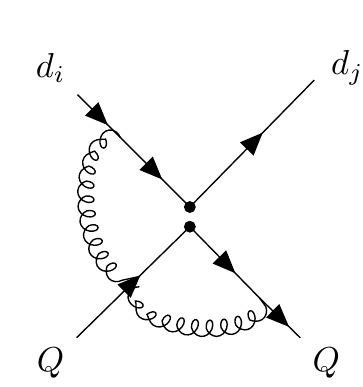}
    \caption{}
    \label{fig:CC-ins-B-1l.pdf}
  \end{subfigure}
\caption{\small
  \label{fig:cc-ins}
  The QCD current-current insertions at one-loop.
}
\end{figure}

As outlined above the transformation of the BMU basis to the JMS basis requires Fierz transformations on some of the BMU operators. This generates then additional contributions to the one-loop  matching performed within the JMS basis. In order to find these
one loop contributions one has to insert the difference $Q_i-\widetilde Q_i$
into current-current and penguin diagrams of Figs.~\ref{fig:cc-ins} and \ref{fig:peng-ins}, respectively.
The details of the calculation for the 1-loop insertions are given in Appendix~\ref{app:master-formulae}.

\begin{figure}[H]
\centering
  \begin{subfigure}[t]{0.32\textwidth}
    \centering
    \includegraphics[trim={5cm 19.1cm 12.9cm 4cm},clip,width=0.7\textwidth]{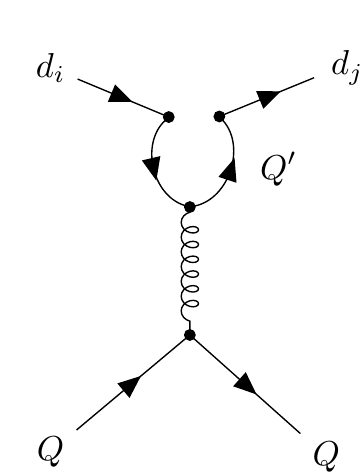}
    \caption{}
    \label{fig:P-ins-open-1l}
  \end{subfigure}
  \begin{subfigure}[t]{0.32\textwidth}
    \centering
    \includegraphics[trim={5cm 19.1cm 12.9cm 4cm},clip,width=0.7\textwidth]{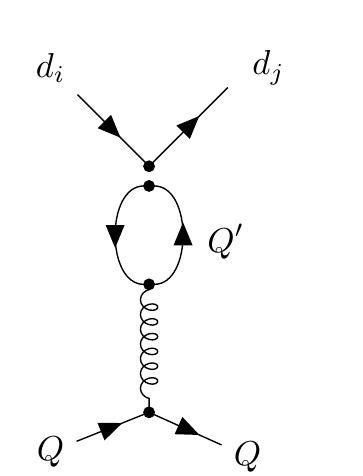}
    \caption{}
    \label{fig:P-ins-closed-1l}
  \end{subfigure}
\caption{\small
  \label{fig:peng-ins}
  The QCD-penguin insertions with
  open--type [left] and closed--type [right] fermion loops.
}
\end{figure}

 With all BMU operators listed in Appendix~\ref{app:def-BMU} it turns out that
\begin{itemize}
\item
  For $\OpL{k}$ with $k=1-18$ these additional contributions come only from penguin
  insertions and moreover only for a few among these operators listed below.
\item
  Fierz transformations on $\OpL[\text{SLR}, Q]{1}$ and $\OpL[\text{SLR}, Q]{2}$
  do not generate any evanescent contributions and consequently in this
  case $D=4$ identities can be used.
\item
  For $\OpL[\text{SRR}, Q]{k}$ with $k=1-4$ and $\OpL[\text{SRR}, D]{l}$ with
  $l=1,2$ the contributions come only from current-current operators and
  involve all operators considered. However, all these operators do not
  contribute to $K$ and $B$ decays being forbidden within SMEFT.
  For completeness we list these contributions below because they could
  be useful for charm physics.
\end{itemize}

In the case of the SM operators $Q_k$ with $k=1-10$ all the
contributions from Fierz transformations for LL (left-left) operators can be obtained from two
properties:
\be
\label{eq:ruleQ1Q2}
  \OpL{1} = \OpLt{1}, \qquad
  \OpL{2} = \OpLt{2} + \frac{1}{3}\frac{\alS}{4\pi} P,
\ee
with
\be
  P = \OpL{4} + \OpL{6} - \frac{1}{3}(\OpL{3} + \OpL{5}).
  \ee
  We find
      \be\label{eq:ruleQ3Q6}
  \OpL{3} = \OpLt{3}+ \frac{2}{3}\frac{\alS}{4\pi} P    , \qquad    
\OpL{4} = \OpLt{4} - \frac{N_f}{3}\frac{\alS}{4\pi} P, \qquad  
\OpL{5} = \OpLt{5}, \qquad  \OpL{6} = \OpLt{6}\,,
\ee
\be\label{eq:ruleQ7Q10}
 \OpL{7} = \OpLt{7}, \qquad  \OpL{8} = \OpLt{8}, \qquad
  \OpL{9} = \OpLt{9}- \frac{1}{3}\frac{\alS}{4\pi} P    , \qquad
  \OpL{10} = \OpLt{10} - \frac{1}{3}(N_u-\frac{N_d}{2})\frac{\alS}{4\pi} P.
  \ee

  We observe that  the Fierz transformations on the VLR operators
  $\OpL{k}$ with $k=5-8$ do not bring any contributions from evanescent
  operators.

In the case of the NP operators $\OpL{k}$ with $k=11-18$ only the Fierz
transformation on  $\OpL{11}$ brings a contribution from evanescent operators
so that
\be \label{eq:ruleQ11}
\OpL{11} = \OpLt{11} +\frac{2}{3}\frac{\alS}{4\pi} P,
\qquad \OpL{k} = \OpLt{k},\quad k=12-18\,.
\ee

The corresponding results for
$\OpL[\text{SRR}, D]{1,2,3,4}$ with $D=d_i$ or $d_j$ operators can be obtained by using the results of
\cite{Buras:2012gm}, in particular the results in equations (29)-(32) of that
paper. In this case the $\OpLt[\text{SRR}, D]{k}$  operators  with $k=1-4$ are
given as follows:
\begin{align}
  \label{FRR1}
  \OpLt[\text{SRR},D]{1} &
  = -\frac{1}{2} \OpL[\text{SRR},D]{2} + \frac{1}{8} \OpL[\text{SRR},D]{4},
&
  \OpLt[\text{SRR},D]{2} &
  = -\frac{1}{2} \OpL[\text{SRR},D]{1} + \frac{1}{8} \OpL[\text{SRR},D]{3},
\\
  \label{FRR2}
  \OpLt[\text{SRR},D]{3} &
  = 6 \OpL[\text{SRR},D]{2} + \frac{1}{2} \OpL[\text{SRR},D]{4},
&
  \OpLt[\text{SRR},D]{4} &
  = 6 \OpL[\text{SRR},D]{1} + \frac{1}{2} \OpL[\text{SRR},D]{3}.
\end{align}

As this time only current-current diagrams are involved the flavour structure
relative to the one considered in \cite{Buras:2012gm} does not matter and the
full calculation of the matrix elements of $\widetilde Q_i$  operators can be
readily performed in no time using results of \cite{Buras:2012gm}.
The shifts caused by evanescent operators involve four operators but those with
$k=1,3$ can be eliminated\footnote{Equivalently one could also eliminate any other two operators but here we follow the
conventions used in Ref.\cite{Aebischer:2021raf}.} using
\begin{align}
  \label{FR1}
  \OpL[\text{SRR}, D]{3} &
  = 6\, \OpL[\text{SRR}, D]{2} + \frac{1}{2}\, \OpL[\text{SRR}, D]{4} + \text{Fierz ev.} 
\\
  \label{FR2}
  \OpL[\text{SRR}, D]{1} &
  = -\frac{1}{2}\, \OpL[\text{SRR}, D]{2} + \frac{1}{8}\, \OpL[\text{SRR}, D]{4} + \text{Fierz ev.} 
\end{align}
so that the shifts depend only on $\OpL[\text{SRR},D]{2,4}$.
{Note that in the BMU basis $\OpL[\text{SRR},D]{2,4}$ are
 denoted as $\OpL[\text{SRR},i]{1,2}$ or $\OpL[\text{SRR},j]{1,2}$ in Ref.\cite{Aebischer:2021raf},
see Appendix \ref{app:def-BMU} for definitions.

We find then
\begin{align}
  \OpL[\text{SRR},D]{1} & = \OpLt[\text{SRR,D}]{1}
  + \frac{\alS}{4\pi} \sum_{k=2,4} A_k \OpL[\text{SRR,D}]{k}, \label{eq:ruleSRRD1}
\\
  \OpL[\text{SRR},D]{2} & = \OpLt[\text{SRR,D}]{2}
  + \frac{\alS}{4\pi} \sum_{k=2,4} B_k \OpL[\text{SRR,D}]{k}, \label{eq:ruleSRRD2}
\\
  \OpL[\text{SRR},D]{3} & = \OpLt[\text{SRR,D}]{3}
  + \frac{\alS}{4\pi} \sum_{k=2,4} C_k \OpL[\text{SRR,D}]{k},
\\
  \OpL[\text{SRR},D]{4} & = \OpLt[\text{SRR,D}]{4}
  + \frac{\alS}{4\pi} \sum_{k=2,4} D_k\OpL[\text{SRR,D}]{k},
\end{align}
with the coefficients $A_k,B_k,C_k,D_k$ given as follows\footnote{$N_c$ is the number of colours with $N_c=3$ in the final results.}

\begin{align}
 A_2 & =\frac{1}{2} + \frac{5}{N_c} - \frac{7 N_c}{4}= -\frac{37}{12},
&
  A_4 & = -\frac{1}{2} + \frac{1}{4 N_c} - \frac{N_c}{16}=-\frac{29}{48}\,.
\end{align}

\begin{align}
  B_2 & = -\frac{17}{4}-\frac{1}{N_c}=-\frac{55}{12},
&
 B_4 & = -\frac{3}{16} +\frac{3}{4N_c}-\frac{N_c}{8}=-\frac{5}{16}\,.
\end{align}

\begin{align}
  C_2 &  =  36 +\frac{28}{N_c} - 7 N_c=\frac{73}{3},
&
 C_4 & = -\frac{1}{2} - \frac{5}{N_c} + \frac{3 N_c}{4}=\frac{1}{12}\,.
\end{align}

\begin{align}
  D_2 & = -21 - \frac{44}{N_c} + 14 N_c=\frac{19}{3},
&
  D_4 &  = \frac{13}{4}+\frac{1}{N_c}=\frac{43}{12}\,.
\end{align}

The same procedure can be applied to $\OpL[\text{SRR},Q]{1,2,3,4}$ with $Q=u_k, d_k \ne d_i, d_j$ defined in eq.~\eqref{eq:def-BMU-SRR-ops}. The rules
for the shifts read

\begin{align}\label{AJB1}
  \OpL[\text{SRR},Q]{1} & = \OpLt[\text{SRR,Q}]{1}
  + \frac{\alS}{4\pi} \sum_{k=1,2,3,4} a_k \OpL[\text{SRR,Q}]{k}\,,
\\
  \OpL[\text{SRR},Q]{2} & = \OpLt[\text{SRR,Q}]{2}
  + \frac{\alS}{4\pi} \sum_{k=1,2,3,4} b_k \OpL[\text{SRR,Q}]{k}\,,\label{AJB2}
\\
  \OpL[\text{SRR},Q]{3} & = \OpLt[\text{SRR,Q}]{3}
  + \frac{\alS}{4\pi} \sum_{k=1,2,3,4} c_k \OpL[\text{SRR,Q}]{k}\,, \label{AJB3}
\\
  \OpL[\text{SRR},Q]{4} & = \OpLt[\text{SRR,Q}]{4}
  + \frac{\alS}{4\pi} \sum_{k=1,2,3,4} d_k\OpL[\text{SRR,Q}]{k}\,. \label{AJB4}
\end{align}
Here the flavour structure of the tilde operators $\OpLt[\text{SRR},Q]{1,2,3,4}$ (see \eqref{FRR1},\eqref{FRR2} for the definition with {$D$} replaced by {$Q$})
is $(\bar d_j \Gamma   Q)( \bar Q \Gamma   d_i)$ and the BMU operators $\OpL[\text{SRR},Q]{1,2,3,4}$ have the form $(\bar d_j \Gamma   d_i)( \bar Q \Gamma  Q)$.

The coefficients $a_k,b_k,c_k,d_k$ are given as follows
\begin{align}
  a_1 & = \frac{N_c}{2}-\frac{1}{N_c} = \frac{7}{6},
&
  a_2 & = \frac{1}{2},
\\
  a_3 & =-\frac{N_c}{4}+\frac{3}{4N_c}= -\frac{1}{2},
&
  a_4 & =-\frac{1}{2} \,.
\end{align}

\begin{align}
  b_1 & = 1,
&
  b_2 & = -\frac{1}{N_c} = -\frac{1}{3},
\\
  b_3 & = -\frac{5}{8},
&
  b_4 & = -\frac{N_c}{8}+\frac{3}{4N_c} = -\frac{1}{8}.
\end{align}

\begin{align}
  c_1 & ={8}N_c -\frac{44}{N_c} = \frac{28}{3},
&
  c_2 & =36,
\\
  c_3 & =-\frac{N_c}{2}+\frac{1}{N_c} = -\frac{7}{6},
&
  c_4 & =-\frac{1}{2}.
\end{align}

\begin{align}
  d_1 & =30,
&
  d_2 & =14{N_c}-\frac{44}{N_c} = \frac{82}{3},
\\
  d_3 & =-1,
&
  d_4 & =\frac{1}{N_c} = \frac{1}{3}\,.
\end{align}

With these rules we can find the matrix $\hat R$ as defined in (\ref{eq:basic}).

{
\subsection{Transformation matrices at one-loop}\label{sec:treeloop}
In this section we present our final result for the transformation matrices between the
BMU and JMS bases at the 1-loop level. The details of the calculation are given in Section~\ref{sec:calc}.
{The calculation can be split into the three disconnected sectors VLL (Vector Left-Left), 
VLR (Vector Left-Right) and SRR (Scalar Right-Right), which denote the $\gamma_\mu P_L\otimes \gamma^\mu P_L$, $\gamma_\mu P_L\otimes \gamma^\mu P_R$ and $ P_R\otimes P_R$ Dirac structures of the involved four-fermi operators, respectively.}
{
For the BMU operator we use the following reference ordering 
\begin{align}
  \text{VLL} & :
  \big\{ \OpL{1},\; \OpL{2},\;\;
         \OpL{3},\; \OpL{4},\; \OpL{9},\; \OpL{10},\;\;
         \OpL{11},\;\; \OpL{14}
  \big\} , \label{eq:BMU.VLL}
\\
  \text{VLR} & :
  \big\{ \OpL{5},\; \OpL{6},\; \OpL{7},\; \OpL{8},\;\;
         \OpL{12},\; \OpL{13}, \;\; \OpL{15}, \ldots , \OpL{24}
  \big\} , \label{eq:BMU.VLR}
\\
  \text{SRR} & :
  \big\{ \OpL{25}, \ldots , \OpL{40}
  \big\}. \label{eq:BMU.SRR}
\end{align}

\indent
For the JMS basis we use the ordering 
\begin{align}
  \text{VLL} & :
  \big\{ \opL[V1,LL]{ud}{11ji},\; \opL[V8,LL]{ud}{11ji},\;
         \opL[V1,LL]{ud}{22ji},\; \opL[V8,LL]{ud}{22ji},
\notag
\\ & \phantom{: \big\{ \;}
         \opL[V,LL]{dd}{jikk}, \; \opL[V,LL]{dd}{jkki},\;
         \opL[V,LL]{dd}{jiii}, \; \opL[V,LL]{dd}{jijj}
  \big\} , \label{eq:JMS.VLL}
\\
  \text{VLR} & :
  \big\{ \opL[V1,LR]{du}{ji11},\; \opL[V8,LR]{du}{ji11},\;
         \opL[V1,LR]{du}{ji22},\; \opL[V8,LR]{du}{ji22},
\notag
\\ & \phantom{: \big\{ \;}
         \opL[V1,LR]{dd}{jikk}, \; \opL[V8,LR]{dd}{jikk},\;
         \opL[V1,LR]{dd}{jiii}, \; \opL[V8,LR]{dd}{jiii},\;
\notag
\\ & \phantom{: \big\{ \;}
         \opL[V1,LR]{dd}{jijj}, \; \opL[V8,LR]{dd}{jijj}, \;
         \opL[V1,LR]{uddu}{1ji1}^\dagger,\; \opL[V8,LR]{uddu}{1ji1}^\dagger, \;
\notag
\\ & \phantom{: \big\{ \;}
         \opL[V1,LR]{uddu}{2ji2}^\dagger,\; \opL[V8,LR]{uddu}{2ji2}^\dagger,
         \opL[V1,LR]{dd}{jkki}, \; \opL[V8,LR]{dd}{jkki}
  \big\} , \label{eq:JMS.VLR}
\\
  \text{SRR} & :
  \big\{ \opL[S1,RR]{dd}{jiii},\; \opL[S8,RR]{dd}{jiii},\;
         \opL[S1,RR]{dd}{jijj},\; \opL[S8,RR]{dd}{jijj},\;
\notag
\\ & \phantom{: \big\{ \;}
         \opL[S1,RR]{ud}{11ji},\;   \opL[S8,RR]{ud}{11ji},\;
         \opL[S1,RR]{uddu}{1ij1},\; \opL[S8,RR]{uddu}{1ij1},
\notag
\\ & \phantom{: \big\{ \;}
         \opL[S1,RR]{ud}{22ji},\;   \opL[S8,RR]{ud}{22ji},\;
         \opL[S1,RR]{uddu}{2ij2},\; \opL[S8,RR]{uddu}{2ij2},
\notag
\\ & \phantom{: \big\{ \;}
         \opL[S1,RR]{dd}{jikk},\; \opL[S8,RR]{dd}{jikk},\;
         \opL[S1,RR]{dd}{jkki},\; \opL[S8,RR]{dd}{jkki}
  \big\} .  \label{eq:JMS.SRR}
\end{align}

}
At tree-level the transformation matrix $\hat R^{(0)}$  reads
\begin{align}
  \hat R^{(0)} & =
  \begin{pmatrix}
    \hat R^{(0)}_\text{VLL} & 0_{8\times 16}    & 0_{8 \times 16} \\
    0_{16\times 8}    & \hat R^{(0)}_\text{VLR} & 0_{16\times 16} \\
    0_{16\times 8}    & 0_{16\times 16}   & \hat R^{(0)}_\text{SRR}
  \end{pmatrix} \,, \label{eq:rmatrix}
\end{align}

\begin{equation}
  \hat R^{(0)}_\text{VLL} =
  \begin{pmatrix}
 1& 0& 0& 0& 0& 0& 0& 0 \\[0.1cm]
 -\frac{1}{6}&  \frac{1}{2}& 0& 0& 0& 0& 0& 0 \\[0.1cm]
 -1& 0& \frac{1}{3}& 0& \frac{2}{3}& 0& 0& 0 \\[0.1cm]
 \frac{1}{6}& -\frac{1}{2}& -\frac{1}{18} & \frac{1}{6}& -\frac{1}{9}& \frac{1}{3}& 0& 0 \\[0.1cm]
 0& 0& \frac{2}{3}& 0& -\frac{2}{3}& 0& -1& 0 \\[0.1cm]
 0& 0& 0& \frac{2}{3}& 0& -\frac{2}{3}& -1& 0 \\[0.1cm]
 0& 0& 0& 0& 0& 0& \frac{1}{2}& \frac{1}{2} \\[0.1cm]
 0& 0& 0& 0& 0& 0& \frac{1}{2}& -\frac{1}{2}
\end{pmatrix} \,,
\qquad
\hat R^{(0)}_\text{SRR} =
  \begin{pmatrix}
  \hat A_\text{SRR}^{(0)} & 0_{2\times 2}     & 0_{2\times 4}     & 0_{2\times 4}     & 0_{2\times 4} \\[0.1cm]
  0_{2\times 2}     & \hat A_\text{SRR}^{(0)} & 0_{2\times 4}     & 0_{2\times 4}     & 0_{2\times 4} \\[0.1cm]
  0_{4\times 2}     & 0_{4\times 2}     & \hat B_\text{SRR}^{(0)} & 0_{4\times 4}     & 0_{4\times 4} \\[0.1cm]
  0_{4\times 2}     & 0_{4\times 2}     & 0_{4\times 4}     & \hat B_\text{SRR}^{(0)} & 0_{4\times 4} \\[0.1cm]
  0_{4\times 2}     & 0_{4\times 2}     & 0_{4\times 4}     & 0_{4\times 4}     & \hat B_\text{SRR}^{(0)}
  \end{pmatrix} \,\,,
\end{equation}

\begin{equation}
  \hat A_\text{SRR}^{(0)} =
  \begin{pmatrix}
  1  & 0 \\[0.1cm]
    -\frac{5}{12} & \frac{1}{16}
  \end{pmatrix} \,,
  \qquad
  \hat B_\text{SRR}^{(0)} =
  \begin{pmatrix}
  0 & 1 & 0 & 0 \\[0.1cm]
  \frac{1}{2} & -\frac{1}{6} & 0 & 0 \\[0.1cm]
  -\frac{1}{2} & 0 & \frac{1}{8}  & 0 \\[0.1cm]
  \frac{1}{12} & -\frac{1}{4} & -\frac{1}{48} & \frac{1}{16}
  \end{pmatrix} \,,
\end{equation}

\bigskip
\bigskip

\begin{equation}\label{eq:R0VLR}
  \hat R^{(0)}_\text{VLR} =
  \begin{pmatrix}
 \frac{1}{6} & 0& \frac{1}{3} & 0& 0& 0& 0& 0& 0& \frac{1}{2}& 0& 0& 0& 0& 0& 0 \\[0.1cm]
 -\frac{1}{36}& \frac{1}{12}& -\frac{1}{18}& \frac{1}{6}& 0& 0 & 0& 0& \frac{1}{4}& -\frac{1}{12}& 0& 0& 0& 0& 0& 0 \\[0.1cm]
 \frac{1}{6}& 0& \frac{1}{3}& 0& 0& 0& 0& 0& 0& -\frac{1}{2}& 0& 0& 0& 0& 0& 0 \\[0.1cm]
 -\frac{1}{36}& \frac{1}{12}& -\frac{1}{18}& \frac{1}{6}& 0& 0& 0& 0& -\frac{1}{4}& \frac{1}{12}& 0& 0& 0& 0&  0& 0 \\[0.1cm]
 \frac{2}{3}& 0& -\frac{2}{3}& 0& 0& -1& 0& 0& 0& 0& 0& 0& 0& 0& 0& 0 \\[0.1cm]
 -\frac{1}{9}& \frac{1}{3}& \frac{1}{9}& -\frac{1}{3}& -\frac{1}{2}& \frac{1}{6}& 0& 0& 0& 0& 0& 0& 0& 0& 0& 0 \\[0.1cm]
 0& 0& 0& 0& 0& \frac{1}{2}& 0& \frac{1}{2}& 0& 0& 0& 0& 0& 0& 0& 0 \\[0.1cm]
 0& 0& 0& 0& \frac{1}{4}& -\frac{1}{12}& \frac{1}{4}& -\frac{1}{12}& 0& 0& 0& 0& 0& 0& 0& 0 \\[0.1cm]
 0& 0& 0& 0& 0& \frac{1}{2}& 0& -\frac{1}{2}& 0& 0& 0& 0& 0& 0& 0& 0 \\[0.1cm]
 0& 0& 0& 0& \frac{1}{4}& -\frac{1}{12}& -\frac{1}{4}& \frac{1}{12}& 0& 0& 0& 0& 0& 0& 0& 0 \\[0.1cm]
 0& 0& 0& 0& 0& 0& 0& 0& 0& 0& -2& 0& 0& 0& 0& 0 \\[0.1cm]
 0& 0& 0& 0& 0& 0& 0& 0& 0& 0& \frac{1}{3}& -1& 0& 0& 0& 0 \\[0.1cm]
 0& 0& 0& 0& 0& 0& 0& 0& 0& 0& 0& 0& -2& 0& 0& 0 \\[0.1cm]
 0& 0& 0& 0& 0& 0& 0& 0& 0& 0& 0& 0& \frac{1}{3}& -1& 0& 0 \\[0.1cm]
 0& 0& 0& 0& 0& 0& 0& 0& 0& 0& 0& 0& 0& 0& -2& 0 \\[0.1cm]
 0& 0& 0& 0& 0& 0& 0& 0& 0& 0& 0& 0& 0& 0& \frac{1}{3}& -1
\end{pmatrix}\,.
\end{equation}
\\[1cm]

At 1-loop level the corrections due to the EV operators are given by the matrix $\hat R^{(1)}$

\begin{align}
  \hat R^{(1)} & =
  \begin{pmatrix}
    \hat R^{(1)}_\text{VLL} & R^{(1)}_{8\times 16}    & 0_{8 \times 16} \\
    0_{16\times 8}    & \hat R^{(1)}_\text{VLR} & 0_{16\times 16} \\
    0_{16\times 8}    & 0_{16\times 16}   & \hat R^{(1)}_\text{SRR}
  \end{pmatrix} \,, \label{eq:R1}
\end{align}

\begin{equation}
  \hat R^{(1)}_\text{VLL} =
  \begin{pmatrix}
  0 & 0 & 0 & 0 & 0 & 0 & 0 & 0 \\[0.1cm]
  0 & 0 & \frac{1}{18} & -\frac{1}{6} & 0 & 0 & 0 & 0 \\[0.1cm]
  0 & 0 & 0 & 0 & 0 & 0 & 0 & 0 \\[0.1cm]
  0 & 0 & -\frac{1}{18} & \frac{1}{6} & 0 & 0 & 0 & 0 \\[0.1cm]
  0 & 0 & 0 & 0 & 0 & 0 & 0 & 0 \\[0.1cm]
  0 & 0 & -\frac{1}{3} & 1 & 0 & 0 & 0& 0 \\[0.1cm]
  0 & 0 & 0 & 0 & 0 & 0 & 0 &  0 \\[0.1cm]
  0 & 0 & 0 & 0 & 0 & 0 & 0 & 0
  \end{pmatrix} \,,
  \qquad
 \hat R^{(1)}_{8 \times 16} =
  \begin{pmatrix}
 0 & 0 &  \dots& 0  \\[0.1cm]
 \frac{1}{18} & -\frac{1}{6} &  \dots& 0  \\[0.1cm]
 0 & 0 &  \dots & 0 \\[0.1cm]
 -\frac{1}{18}& \frac{1}{6}& \dots &  0 \\[0.1cm]
 0 & 0 & \dots &0 \\[0.1cm]
 -\frac{1}{3}& 1& \dots&  0 \\[0.1cm]
 0 & 0 & \dots & 0 \\[0.1cm]
 0 & 0 & \dots & 0
\end{pmatrix} \,,
\qquad
\hat R^{(1)}_\text{VLR} = 0_{16\times 16}\,,
\end{equation}

\begin{equation}
  \hat R^{(1)}_\text{SRR} =
  \begin{pmatrix}
  \hat A^{(1)}_\text{SRR} & 0_{2\times 2}     & 0_{2\times 4}     & 0_{2\times 4}     & 0_{2\times 4} \\[0.1cm]
  0_{2\times 2}     & \hat A^{(1)}_\text{SRR} & 0_{2\times 4}     & 0_{2\times 4}     & 0_{2\times 4} \\[0.1cm]
  0_{4\times 2}     & 0_{4\times 2}     & \hat B^{(1)}_\text{SRR} & 0_{4\times 4}     & 0_{4\times 4} \\[0.1cm]
  0_{4\times 2}     & 0_{4\times 2}     & 0_{4\times 4}     & \hat B^{(1)}_\text{SRR} & 0_{4\times 4} \\[0.1cm]
  0_{4\times 2}     & 0_{4\times 2}     & 0_{4\times 4}     & 0_{4\times 4}     & \hat B^{(1)}_\text{SRR}
  \end{pmatrix} \,,
\quad
  \begin{array}{c}
  \begingroup
  \setlength\arraycolsep{0.15cm}
  \hat A^{(1)}_\text{SRR} =
  \begin{pmatrix}
    0 & 0 \\[0.1cm]
    -\frac{37}{24} & -\frac{29}{96}
  \end{pmatrix} \,,
  \endgroup
  \\ \quad
  \hat B^{(1)}_\text{SRR} =
  \begin{pmatrix}
  0 & 0 & 0 & 0 \\[0.1cm]
  0 & 0 & 0 & 0 \\[0.1cm]
  -\frac{7}{12} & -\frac{17}{4} & -\frac{5}{48} & -\frac{3}{16} \\[0.1cm]
  -\frac{55}{36} & -\frac{13}{12} & -\frac{11}{144} & -\frac{1}{48}
  \end{pmatrix} \,.
  \end{array}\label{eq:R1SRR}
\end{equation}
}

{Here, we have used $N_f=5, N_d=3$ and $N_u=2$.}

\section{Evanescent operators}
\label{sec:calc}
In this section, we define and calculate the EV operators which in turn gives us the
matrix $\hat R^{(1)}$.

\subsection{{Definition of EV operators in the} VLL sector}

{At the 1-loop level, we need to add an EV contribution to a tree-level basis transformation rule if
it involves a Fierz transformation. On the other hand, we do not need an EV contribution
for the cases in which no Fierz is required for the tree-level basis change. In the VLL sector, there are total
eight independent operators where only five of them involve Fierz relation \eqref{eq:fierzvll} for the change
of basis. The corresponding EV operators $\EV[VLL]{I}, I=1-5$ are defined by the relations:}
\begin{eqnarray}
\opL[V1,LL]{ud}{11ji} & \stackrel{\mathcal{F}}{=} & \OpL{1} + \EV[VLL]{1}\,,  \label{eq:V1LLud11ji}  \\
\opL[V8,LL]{ud}{11ji} & \stackrel{\mathcal{F}}{=} & -\frac{1}{6}\OpL{1} + \frac{1}{2} \OpL{2} +\EV[VLL]{2}\,, \label{eq:V8LLud11ji}   \\
\opL[V1,LL]{ud}{22ji} & \stackrel{\mathcal{F}}{=} & -\OpL{1}  +\frac{1}{3} \OpL{3} +\frac{2}{3} \OpL{9} +\EV[VLL]{3}\,, \label{eq:V1LLud22ji}  \\
\opL[V8,LL]{ud}{22ji} & \stackrel{\mathcal{F}}{=} & \frac{1}{6} \OpL{1}  -\frac{1}{2} \OpL{2} -\frac{1}{18} \OpL{3}
+ \frac{1}{6} \OpL{4} -\frac{1}{9} \OpL{9} + \frac{1}{3} \OpL{10} +\EV[VLL]{4} \,,  \label{eq:V8LLud22ji} \\
\opL[V,LL]{dd}{jikk} & = & \frac{2}{3}\OpL{3}  -\frac{2}{3} \OpL{9} - \OpL{11}\,, \label{eq:VLLddjikk}   \\
\opL[V,LL]{dd}{jkki} & \stackrel{\mathcal{F}}{=} & \frac{2}{3} \OpL{4} -\frac{2}{3} Q_{10} - \OpL{11} + \EV[VLL]{5}\,, \label{eq:VLLddjkki}   \\
\opL[V,LL]{dd}{jiii} & = & \frac{1}{2}\OpL{11} +\frac{1}{2} \OpL{14}  \label{eq:VLLddjiii}   \,,   \\
\opL[V,LL]{dd}{jijj} & = & \frac{1}{2}\OpL{11} - \frac{1}{2} \OpL{14}.  \label{eq:VLLddjijj}
\end{eqnarray}
Here $\mathcal{F}$ indicates that the Fierz identity \eqref{eq:fierzvll} is needed for the
change of basis.
{
\subsection{Calculation of the EV operators in the VLL sectors}
}

Now we are in position to use the rules presented in Sec.~\ref{subsec:brules} to obtain
$\EV[VLL]{1}$ -$\EV[VLL]{5}$ which contribute to $\hat R^{(1)}$.
{In order to use the rules of Sec.\ref{subsec:brules} first we need to express the JMS operators on the l.h.s. in terms of the
$\OpLt{I}$ operators.} {In general}, there are three categories of operators as discussed in Sec.~\ref{subsec:procedure}.

\subsubsection{Operators requiring no Fierz}
Since the following set of operators in the VLL sector do not require Fierz
transformations for the basis change
\begin{equation}
\opL[V,LL]{dd}{jikk}\,, \opL[V,LL]{dd}{jiii}\,, \opL[V,LL]{dd}{jijj}\,,
\end{equation}
there are no EV operator contributions at the 1-loop level {basis transformation given by
\eqref{eq:VLLddjikk},\eqref{eq:VLLddjiii} and \eqref{eq:VLLddjijj}.}
Hence, the corresponding entries in the matrix $\hat R^{(1)}$ vanish.

\subsubsection{Operators requiring Fierz but no EV shifts}
 {There are two operators in the JMS basis which require Fierz transformation
but the EV contribution still vanish. The tree-level transformations for these operators
 are given by \eqref{eq:V1LLud11ji} and \eqref{eq:V1LLud22ji}. To see this, one has to express
the JMS operators on the LHS in terms of the $\OpLt{I}$ operators, doing so we obtain}
\begin{eqnarray}
\EV[VLL]{1}  &=& \opL[V1,LL]{ud}{11ji} - \OpL{1}    ~= ~ \OpLt{1}-\OpL{1}  =0\,, \\
\EV[VLL]{3}  &=& \opL[V1,LL]{ud}{22ji} - \left ( -\OpL{1} + \frac{1}{3} \OpL{3}
+\frac{2}{3} \OpL{9}  \right )  ~= ~ \OpLt{1}-\OpL{1} = ~0.
\end{eqnarray}
{Here the shift in $\OpLt{1} - \OpL{1}$ is given by the rule \eqref{eq:ruleQ1Q2}.}

\subsubsection{Operators requiring Fierz and EV shifts}
Finally, we turn to the cases for which Fierz transformation at the tree-level
as well as the EV contributions at the 1-loop level are necessary for the basis
transformation. The tree-level transformations can be read from Eqs.\eqref{eq:V8LLud11ji}, \eqref{eq:V8LLud22ji}
and \eqref{eq:VLLddjkki}. The EV operators are then given by
\begin{eqnarray}
\EV[VLL]{2} &=&  \opL[V8,LL]{ud}{11ji} - \left( -\frac{1}{6} \OpL{1} +\frac{1}{2} \OpL{2}
\right )  \nn \\
  &=&\frac{1}{6} (\OpL{1}-\OpLt{1}) - \frac{1}{2}  (\OpL{2}-\OpLt{2})
  ~=~  -\frac{1}{6} \frac{\alS}{4\pi} P\,, \\
\EV[VLL]{4}  &=& = \opL[V8,LL]{ud}{22ji} -
\left ( \frac{1}{6} \OpL{1} -\frac{1}{2} \OpL{2} -\frac{1}{18} \OpL{3}
+\frac{1}{6} \OpL{4} -\frac{1}{9} \OpL{9}  +\frac{1}{3} \OpL{10}  \right )            \nn \\
&=&  \frac{1}{6}(\OpLt{1}-\OpL{1})+\frac{1}{2}(\OpL{2}-\OpLt{2})=\frac{1}{6} \frac{\alS}{4\pi} P \,,\\
\EV[VLL]{5}  &=& \opL[V,LL]{dd}{jkki}  -
\left ( \frac{2}{3} \OpL{4} -\frac{2}{3} \OpL{10} - \OpL{11} \right )  \nn\\
&=&{\frac{2}{3} (\OpLt{4}-\OpL{4}) -\frac{2}{3} (\OpLt{10}-\OpL{10}) }\nn\\
&=&{\frac{2N_f+N_d-2N_u}{9} \frac{\alS}{4\pi} P}.
\end{eqnarray}

{Here the 1-loop shifts are given by the rules in \eqref{eq:ruleQ1Q2},\eqref{eq:ruleQ3Q6},\eqref{eq:ruleQ7Q10} and
\eqref{eq:ruleQ11}, respectively. }

\subsection{Definition of EV operators in the SRR sector}
{In this case, in addition to the color identity \eqref{eq:colorid}, we need the Fierz relations given in
\eqref{eq:fierzSRR} and \eqref{eq:fierzTRR} \footnote{Note that here we have used the definition
$\sigma_{\mu\nu} = \frac{i}{2}[\gamma_\mu, \gamma_\nu]$. Also we define the operators $\OpL[\text{SRR},i]{2}$ and
$\OpL[\text{SRR},Q]{3}$,   $\OpL[\text{SRR},Q]{4}$ with an additional negative sign as compared to
Ref.\cite{Aebischer:2021raf}. } }
The 1-loop basis transformations {including the EV operators $E_I^{\text{SRR}}$ read}
\begin{eqnarray}
\opL[S1,RR]{dd}{jiii} &  = & {\OpL[\text{SRR},i]{2}}\,,   \\
\opL[S8,RR]{dd}{jiii} & \stackrel{\mathcal{F}}{=} & -\frac{5}{12} {\OpL[\text{SRR},i]{2}} +\frac{1}{16}  {\OpL[\text{SRR},i]{4}}
+ \EV[SRR,i]{}. \label{eq:s8RRjiii}
\end{eqnarray}
Similar relations hold for $jiii \to jijj$.  We note that a separate evanescent operator
$\EV[SRR,j]{}$ is needed for this relation.

For the SRR,Q operators we find:}

\begin{eqnarray}
\opL[S1,RR]{ud}{11ji} & = & \OpL[\text{SRR},u]{2}  \,,   \\
\opL[S8,RR]{ud}{11ji} & = &  \frac{1}{2}\OpL[\text{SRR},u]{1} -\frac{1}{6} \OpL[\text{SRR},u]{2}\,,   \\
\opL[S1,RR]{uddu}{1ij1} & \stackrel{\mathcal{F}}{=} & -\frac{1}{2} \OpL[\text{SRR},u]{1} +\frac{1}{8} \OpL[\text{SRR},u]{3}
+ \EV[SRR,u]{1} \,, \label{eq:s1RRuddu1ij1}  \\
\opL[S8,RR]{uddu}{1ij1} & \stackrel{\mathcal{F}}{=} & \frac{1}{12}\OpL[\text{SRR},u]{1} - \frac{1}{4} \OpL[\text{SRR},u]{2}
-\frac{1}{48} \OpL[\text{SRR},u]{3} + \frac{1}{16} \OpL[\text{SRR},u]{4}+ \EV[SRR,u]{2}. \label{eq:s8RRuddu1ij1}
\end{eqnarray}
Similar relations hold for $1 \to 2$ and $u\to c$ on the l.h.s. and r.h.s., respectively. Finally,
\begin{eqnarray}
\opL[S1,RR]{dd}{jikk} & = &  \OpL[\text{SRR},d_k]{2} \,,\\
\opL[S8,RR]{dd}{jikk} & =  &  \frac{1}{2}\OpL[\text{SRR},d_k]{1} - \frac{1}{6}\OpL[\text{SRR},d_k]{2}
 \,,\\
\opL[S1,RR]{dd}{jkki} & \stackrel{\mathcal{F}}{=} &
-\frac{1}{2} \OpL[\text{SRR},d_k]{1} +\frac{1}{8} \OpL[\text{SRR},d_k]{3} +\EV[SRR,$d_k$]{1} \,,  \label{eq:s1RRuddujkki}  \\
\opL[S8,RR]{dd}{jkki} & \stackrel{\mathcal{F}}{=} & \frac{1}{12}\OpL[\text{SRR},d_k]{1}
- \frac{1}{4} \OpL[\text{SRR},d_k]{2} - \frac{1}{48} \OpL[\text{SRR},d_k]{3}
+ \frac{1}{16} \OpL[\text{SRR},d_k]{4} + \EV[SRR,$d_k$]{2}.  \label{eq:s8RRuddujkki}
\end{eqnarray}

{
\subsection{Calculation of EV operators in the SRR sector}
}
\subsubsection{Operators requiring no Fierz}
In the SRR sector, the following operators do not require Fierz transformations for the basis change
at the tree-level
\begin{align}
\opL[S1,RR]{dd}{jiii}\,,&{\opL[S1,RR]{dd}{jijj}\,,} \opL[S1,RR]{ud}{11ji}\,, \opL[S8,RR]{ud}{11ji} \,, {\opL[S1,RR]{ud}{22ji}\,, }\\\notag
&{\opL[S8,RR]{ud}{22ji} \,,}\opL[S1,RR]{dd}{jikk}\,,  \opL[S8,RR]{ud}{jikk}.
\end{align}
Therefore, for the basis change at the 1-loop level no EV contributions are required.

\subsubsection{Operators requiring Fierz but no EV shifts}
{In this sector there are no such operators which require Fierz without having non vanishing EV shifts at
the 1-loop level. }

\subsubsection{Operators requiring Fierz and EV shifts}
The SRR operators which require the Fierz relation and non vanishing EV operators for the
basis transformation are given by \eqref{eq:s8RRjiii}, \eqref{eq:s1RRuddu1ij1}, \eqref{eq:s8RRuddu1ij1}, \eqref{eq:s1RRuddujkki},
and \eqref{eq:s8RRuddujkki}. The EV operators are then given by

\begin{eqnarray}
\EV[SRR,i]{} & = & \opL[S8,RR]{dd}{jiii} - \left ( -\frac{5}{12} \OpL[\text{SRR},i]{2} + {\frac{1}{16}} \OpL[\text{SRR},i]{4 }
 \right )  = {-}\frac{1}{2}(\OpLt[\text{SRR},i]{1}-\OpL[\text{SRR},i]{1}) \nn\\
&=& {-}\frac{\alpha_s}{4\pi}\left(\frac{37}{24}\OpL[\text{SRR},i]{2}+\frac{29}{96}\OpL[\text{SRR},i]{4}\right) \,, \\
\EV[SRR,u]{1}  &=& \opL[S1,RR]{uddu}{1ij1} - \left (  -\frac{1}{2} \OpL[\text{SRR},u]{1} +\frac{1}{8} \OpL[\text{SRR},u]{3}
 \right ) \nn\\
  &=& -\frac{1}{2} (\OpLt[\text{SRR},u]{1}-\OpL[\text{SRR},u]{1}) +\frac{1}{8} (\OpLt[\text{SRR},u]{3}-\OpL[\text{SRR},u]{3})\nn\\
 &=&\frac{\alpha_s}{4\pi} \sum_{k=1,2,3,4} p_k \OpL[\text{SRR,u}]{k}\,, \\
\EV[SRR,u]{2} &=&  \opL[S8,RR]{uddu}{1ij1} -\left ( \frac{1}{12}\OpL[\text{SRR},u]{1} - \frac{1}{4} \OpL[\text{SRR},u]{2}
-\frac{1}{48} \OpL[\text{SRR},u]{3} + \frac{1}{16} \OpL[\text{SRR},u]{4} \right )  \nn \\
&=& \frac{1}{12}(\OpLt[\text{SRR},u]{1}-\OpL[\text{SRR},u]{1}) - \frac{1}{4} (\OpLt[\text{SRR},u]{2}-\OpL[\text{SRR},u]{2})\nn \\
& -&   \frac{1}{48} (\OpLt[\text{SRR},u]{3}-\OpL[\text{SRR},u]{3}) 
+ \frac{1}{16} (\OpLt[\text{SRR},u]{4}-\OpL[\text{SRR},u]{4}) \nn\\
&=&  \frac{\alpha_s}{4\pi} \sum_{k=1,2,3,4} q_k \OpL[\text{SRR,u}]{k}.
\end{eqnarray}
The coefficients $p_k$ and $q_k$ are found to be
\begin{eqnarray}
p_1 &= & - \frac{7}{12} \,, \,\  p_2= - \frac{17}{4} \,, \,\  p_3= -\frac{5}{48} \,, \,\ p_4= -\frac{3}{16} \,,\\
q_1 &= & -\frac{55}{36} \,,  \,\ q_2= -\frac{13}{12} \,, \,\  q_3= -\frac{11}{144} \,,  \,\  q_4= -\frac{1}{48}\,.
\end{eqnarray}

The evanescent operators
$\EV[SRR,j]{}$ and $\EV[SRR,c]{1,2}\,,\EV[SRR,$d_k$]{1,2}$ follow from the corresponding $\EV[SRR,i]{}$ and $\EV[SRR,u]{1,2}$ by the corresponding flavour replacements.

\vspace{0.3cm}
In the above calculation, the shifts $\OpL[\text{SRR},D]{1} - \OpLt[\text{SRR},D]{1} $,   
and {$\OpL[\text{SRR},Q]{I} - \OpLt[\text{SRR},Q]{I} $} for {$Q= u $ or $d_k$} are obtained using the
rules \eqref{eq:ruleSRRD1} and \eqref{AJB1} - \eqref{AJB4},
respectively.  It is worth noting at the 1-loop QCD only current-current insertions are involved for the SRR operators
in obtaining these rules. Therefore flavour structure of the operators is immaterial.
For instance, the operators $\opL[S1,RR]{uddu}{1ij1}$ and $\OpL[\text{SRR,u}]{2}$
having same color and Lorentz structures can be treated on the same footing
even though they have different flavour structures.

\subsection{Definition of EV operators in the VLR sector}
In this subsection we turn our attention to the VLR sector and define the corresponding evanescent operators. Using the Fierz relation in eq.~\eqref{eq:fierzVLR} as well as colour relations one finds:
\begin{eqnarray}
\opL[V1,LR]{du}{ji11} & \stackrel{\mathcal{F}}{=} & \frac{1}{6} \OpL{5}
+\frac{1}{3} \OpL{7} + \frac{1}{2} \OpL{18} + \EV[VLR]{1}\,, \nn  \\
\opL[V8,LR]{du}{ji11} & \stackrel{\mathcal{F}}{=} &
-\frac{1}{36} \OpL{5} +\frac{1}{12} \OpL{6}  - \frac{1}{18} \OpL{7} +\frac{1}{6} \OpL{8} + \frac{1}{4}\OpL{17} - \frac{1}{12} \OpL{18} +\EV[VLR]{2}\,, \nn  \\
\opL[V1,LR]{du}{ji22} & \stackrel{\mathcal{F}}{=} & \frac{1}{6} \OpL{5}
+\frac{1}{3} \OpL{7} -\frac{1}{2} \OpL{18} +\EV[VLR]{3}\,, \nn  \\
\opL[V8,LR]{du}{ji22} & \stackrel{\mathcal{F}}{=} &
-\frac{1}{36} \OpL{5} + \frac{1}{12} \OpL{6} -\frac{1}{18} \OpL{7} + \frac{1}{6} \OpL{8} -\frac{1}{4} \OpL{17}  +\frac{1}{12} \OpL{18} +\EV[VLR]{4} \,,\nn   \\
\opL[V1,LR]{dd}{jikk} & = & \frac{2}{3} \OpL{5} -\frac{2}{3} \OpL{7} -\OpL{13}\,, \nn  \\
\opL[V8,LR]{dd}{jikk} & \stackrel{\mathcal{F}}{=} &
-\frac{1}{9} \OpL{5} +\frac{1}{3}\OpL{6}
+ \frac{1}{9}\OpL{7}  -\frac{1}{3}\OpL{8} -\frac{1}{2} \OpL{12} +\frac{1}{6} Q_{13}  +\EV[VLR]{5}\,, \nn  \\
\opL[V1,LR]{dd}{jiii} & = & \frac{1}{2}\OpL{13} +\frac{1}{2} \OpL{16}  \,, \nn  \\
\opL[V8,LR]{dd}{jiii} & = & \frac{1}{4}\OpL{12} - \frac{1}{12} \OpL{13}+\frac{1}{4}\OpL{15}
-\frac{1}{12}\OpL{16}\,, \nn \\
\opL[V1,LR]{dd}{jijj} & = & \frac{1}{2}\OpL{13} - \frac{1}{2} \OpL{16}\,, \nn \\
\opL[V8,LR]{dd}{jijj} & = & \frac{1}{4}\OpL{12} - \frac{1}{12} \OpL{13}
-\frac{1}{4}\OpL{15}+\frac{1}{12}\OpL{16}\,, \nn \\
\opL[V1,LR]{uddu}{1ji1}^\dagger & = & -2\OpL{19} \,, \nn \\
\opL[V8,LR]{uddu}{1ji1}^\dagger & = & \frac{1}{3}\OpL{19} - \OpL{20}\,,\nn \\
\opL[V1,LR]{uddu}{2ji2}^\dagger & = & -2\OpL{21} \,,\nn \\
\opL[V8,LR]{uddu}{2ji2}^\dagger & = & \frac{1}{3}\OpL{21} -  \OpL{22}\,,\nn \\
\opL[V1,LR]{dd}{jkki} & = & -2\OpL{23} \,,\nn \\
\opL[V8,LR]{dd}{jkki} & = & \frac{1}{3}\OpL{23} -  \OpL{24}.
\end{eqnarray}

In the VLR sector, the following set operators do not require Fierz transformations for the basis change
at the tree-level
\begin{align}
\opL[V1,LR]{dd}{jikk}&\,,\opL[V1,LR]{dd}{jiii}\,,\opL[V8,LR]{dd}{jiii}\,,\opL[V1,LR]{dd}{jijj}\,,\opL[V8,LR]{dd}{jijj} \,,\opL[V1,LR]{uddu}{1ji1}^\dagger \,, \nn\\
\opL[V8,LR]{uddu}{1ji1}^\dagger &\,, \opL[V1,LR]{uddu}{2ji2}^\dagger \,, \opL[V8,LR]{uddu}{2ji2}^\dagger \,,\opL[V1,LR]{dd}{jkki} \,, \opL[V8,LR]{dd}{jkki}\,.
\end{align}
Therefore, for the basis change at the 1-loop level no EV contributions are required.
The rest of the operators in the VLR sector requiring Fierz are
\begin{equation}
  \opL[V1,LR]{du}{ji11}\,,
  \opL[V8,LR]{du}{ji11}\,,
  \opL[V1,LR]{du}{ji22}\,,
  \opL[V8,LR]{du}{ji22} \,,
  \opL[V8,LR]{dd}{jikk}\,.
\end{equation}
However, as discussed in Subsec.~\ref{BasicM} the corresponding EV vanish:
\begin{equation}
  \EV[VLR]{1}=\EV[VLR]{2}=\EV[VLR]{3}=\EV[VLR]{4}=\EV[VLR]{5}=0.
\end{equation}
%
%
%
\section{\boldmath Conclusions
  \label{sec:conclusion}
}
We have presented a simple recipe to perform one-loop basis transformations involving evanescent operators. {The procedure consists of computing the commutator of a one-loop $(L)$ correction using dimensional regularisation and a Fierz ($\mathcal{F}$) transformation of a given operator $Q$, which in all generality is non-vanishing:

\begin{equation}
  [L,\mathcal{F}]Q \neq 0\,.
\end{equation}

{ The presented method has been already used successfully in several contexts such as NLO basis transformations \cite{Aebischer:2020dsw,Aebischer:2021raf,Aebischer:2021hws}, one-loop matching calculations \cite{Aebischer:2018acj} as well as in several two-loop calculations \cite{Buras:2000if}. But it has
  not been presented in any detail and in particular in this generality
  in the literature so far. The present paper}
should help to clarify possible issues involving evanescent operators. In the coming years one-loop matching and two-loop running effects will become more important in NP  analyses than they are now.

We illustrated the outlined procedure by computing explicitly the complete one-loop basis change from the BMU to the JMS basis at $\mathcal{O}(\alpha_s)$ {and this example should allow the reader to perform the transformation
  between different bases.} In this context our method will serve as a simple tool to perform one-loop basis transformations. One particular example would be the basis change to the CMM (Chetyrkin, Misiak and Munz)  basis \cite{Chetyrkin:1997gb}, which is most suited for multi-loop computations.

{Since the one-loop basis change consists of a series of simple algebraic manipulations, it would be interesting to automate this procedure. After having computed all one-loop corrections to the operators in question a simple algorithm might be included in codes such as for example \texttt{abc-eft} \cite{Proceedings:2019rnh}.}
%
%
%
\section*{Acknowledgments}
J.\,A.\ acknowledges financial support from the European Research Council (ERC) under the European Union’s Horizon 2020 research and innovation programme under grant agreement 833280 (FLAY), and from the Swiss National Science Foundation (SNF) under contract 200020-204428.
A.J.B acknowledges financial support
from the Excellence Cluster ORIGINS,
funded by the Deutsche Forschungsgemeinschaft (DFG, German Research Foundation) under Germany´s Excellence Strategy – EXC-2094 – 390783311.
J.K. is financially supported by the Alexander von Humboldt Foundation's
postdoctoral research fellowship.
%
%
%

\appendix

%
%
%
\section[\boldmath $\Delta F=1$ BMU basis for $\Nf = 5$]
{\bm $\Delta F=1$ BMU basis for $\Nf = 5$}
\label{app:def-BMU}
{In this Appendix we collect the full set of BMU operators.

 We start the list with the vector operators, where the first ten operators are the well known SM operators $Q_1-Q_{10}$:}

\begin{equation}
\begin{aligned}
  \OpL{1} &
  = \OpL[\text{VLL},u]{1}
  = (\bar d_j^\alpha \gamma_\mu P_L u^\beta)
    (\bar u^\beta    \gamma^\mu P_L d_i^\alpha) ,
\\
  \OpL{2} &
  = \OpL[\text{VLL},u]{2}
  = (\bar d_j^\alpha \gamma_\mu P_L u^\alpha)
    (\bar u^\beta    \gamma^\mu P_L d_i^\beta) ,
\end{aligned}
\end{equation}
\begin{equation}
  \label{eq:QCD-peng-op}
\begin{aligned}
  \OpL{3} & = (\bar d_j^\alpha \gamma_\mu P_L d_i^\alpha)
    \!\sum_{q}(\bar q^\beta    \gamma^\mu P_L \, q^\beta) , \qquad
&
  \OpL{4} & = (\bar d_j^\alpha \gamma_\mu P_L d_i^\beta)
    \!\sum_{q}(\bar q^\beta    \gamma^\mu P_L \, q^\alpha) ,
\\
  \OpL{5} & = (\bar d_j^\alpha \gamma_\mu P_L d_i^\alpha)
    \!\sum_{q}(\bar q^\beta    \gamma^\mu P_R \, q^\beta) ,
&
  \OpL{6} & = (\bar d_j^\alpha \gamma_\mu P_L d_i^\beta)
    \!\sum_{q}(\bar q^\beta    \gamma^\mu P_R \, q^\alpha) ,
\end{aligned}
\end{equation}
\begin{equation}
  \label{eq:QED-peng-op}
\begin{aligned}
  \OpL{7} & = \frac{3}{2}\,(\bar d_j^\alpha \gamma_\mu P_L d_i^\alpha)
      \!\sum_{q} \! Q_q \, (\bar q^\beta \gamma^\mu P_R \, q^\beta) ,  \qquad
&
  \OpL{8} & = \frac{3}{2}\,(\bar d_j^\alpha \gamma_\mu P_L d_i^\beta)
      \!\sum_{q} \! Q_q \, (\bar q^\beta \gamma^\mu P_R \, q^\alpha) ,
\\
  \OpL{9} & = \frac{3}{2}\,(\bar d_j^\alpha \gamma_\mu P_L d_i^\alpha)
      \!\sum_{q} \! Q_q \, (\bar q^\beta \gamma^\mu P_L \, q^\beta) ,
&
  \OpL{10} & =\frac{3}{2}\,(\bar d_j^\alpha \gamma_\mu P_L d_i^\beta)
      \!\sum_{q} \! Q_q \, (\bar q^\beta \gamma^\mu P_L \, q^\alpha) .
\end{aligned}
\end{equation}

{The NP (New Physics) vector operators in the BMU basis are given by:}

\begin{equation}
\begin{aligned}
  \OpL{11} = \; \OpL[\text{VLL},i+j]{1} &
  = (\bar d_j^\alpha \gamma_\mu P_L d_i^\alpha) \, \big[
    (\bar d_i^\beta \gamma^\mu P_L d_i^\beta)
  + (\bar d_j^\beta \gamma^\mu P_L d_j^\beta)\big] ,
\\
  \OpL{12} = \; \OpL[\text{VLR},i+j]{1} &
  = (\bar d_j^\alpha \gamma_\mu P_L d_i^\beta) \, \big[
    (\bar d_i^\beta \gamma^\mu P_R d_i^\alpha)
  + (\bar d_j^\beta \gamma^\mu P_R d_j^\alpha)\big] ,
\\
  \OpL{13} = \; \OpL[\text{VLR},i+j]{2} &
  = (\bar d_j^\alpha \gamma_\mu P_L d_i^\alpha) \, \big[
    (\bar d_i^\beta \gamma^\mu P_R d_i^\beta)
  + (\bar d_j^\beta \gamma^\mu P_R d_j^\beta)\big] ,
\end{aligned}
\end{equation}
\begin{equation}
\begin{aligned}
  \OpL{14} = \; \OpL[\text{VLL},i-j]{1} &
  = (\bar d_j^\alpha \gamma_\mu P_L d_i^\alpha) \,\big[
    (\bar d_i^\beta \gamma^\mu P_L d_i^\beta)
  - (\bar d_j^\beta \gamma^\mu P_L d_j^\beta)\big] ,
\\
  \OpL{15} = \; \OpL[\text{VLR},i-j]{1} &
  = (\bar d_j^\alpha \gamma_\mu P_L d_i^\beta) \, \big[
    (\bar d_i^\beta \gamma^\mu P_R d_i^\alpha)
  - (\bar d_j^\beta \gamma^\mu P_R d_j^\alpha)\big] ,
\\
  \OpL{16} = \; \OpL[\text{VLR},i-j]{2} &
  = (\bar d_j^\alpha \gamma_\mu P_L d_i^\alpha) \, \big[
    (\bar d_i^\beta \gamma^\mu P_R d_i^\beta)
  - (\bar d_j^\beta \gamma^\mu P_R d_j^\beta)\big] ,
\end{aligned}
\end{equation}
\begin{equation}
\begin{aligned}
  \OpL{17} = \; \OpL[\text{VLR}, u-c]{1} &
  = (\bar d_j^\alpha \gamma_\mu P_L d_i^\beta) \, \big[
    (\bar u^\beta \gamma^\mu P_R \, u^\alpha)
  - (\bar c^\beta \gamma^\mu P_R \, c^\alpha) \big],
\\
  \OpL{18} = \; \OpL[\text{VLR}, u-c]{2} &
  = (\bar d_j^\alpha \gamma_\mu P_L d_i^\alpha) \, \big[
    (\bar u^\beta \gamma^\mu P_R \, u^\beta)
  - (\bar c^\beta \gamma^\mu P_R \, c^\beta) \big].
\end{aligned}
\end{equation}

{Finally, we introduce the scalar sector of the BMU basis. In the SRL sector we use the structures}
\begin{align}
  \OpL[\text{SRL}, Q]{1} &
  = (\bar d_j^\alpha P_R d_i^\beta) \, (\bar Q^\beta P_L \, Q^\alpha) ,
&
  \OpL[\text{SRL}, Q]{2} &
  = (\bar d_j^\alpha P_R d_i^\alpha) \, (\bar Q^\beta P_L \, Q^\beta) ,
\end{align}

{which define the operators}
\begin{equation}
\begin{aligned}
  (\OpL{19},\, \OpL{20}) & = (\OpL[\text{SRL}, u]{1},\, \OpL[\text{SRL}, u]{2}) ,
  \qquad \qquad
&
\\
  (\OpL{21},\, \OpL{22}) & = (\OpL[\text{SRL}, c]{1},\, \OpL[\text{SRL}, c]{2}) ,
  \qquad \qquad
&
  (\OpL{23},\, \OpL{24}) & = (\OpL[\text{SRL}, d_k]{1},\, \OpL[\text{SRL}, d_k]{2}) .
\end{aligned}
\end{equation}
{ In the SRR sector with three equal quarks we introduce
for completeness the redundant structures
\begin{align}
\OpL[\text{SRR}, i]{1} &
= (\bar d_j^\alpha P_R \, d_i^\beta) \, (\bar d_i^\beta P_R \, d_i^\alpha) ,
&
\OpL[\text{SRR}, i]{3} &
= {-} (\bar d_j^\alpha \sigma_{\mu\nu} P_R \, d_i^\beta) \,
  (\bar d_i^\beta  \sigma^{\mu\nu} P_R \, d_i^\alpha) ,
\end{align}
together with the operators
}
\begin{align}
  \label{eq:BMU-SRR,i}
  \OpL{25} = {\OpL[\text{SRR}, i]{2}} &
  = (\bar d_j^\alpha P_R \, d_i^\alpha) \, (\bar d_i^\beta P_R \, d_i^\beta) ,
&
  \OpL{26} = {\OpL[\text{SRR}, i]{4}} &
  = {-} (\bar d_j^\alpha \sigma_{\mu\nu} P_R \, d_i^\alpha) \,
    (\bar d_i^\beta  \sigma^{\mu\nu} P_R \, d_i^\beta) ,
\intertext{and similar for the SRR$,j$ sector}
  \OpL{27} = {\OpL[\text{SRR}, j]{2}} &
  = (\bar d_j^\alpha P_R \, d_i^\alpha) \, (\bar d_j^\beta P_R \, d_j^\beta) ,
&
  \OpL{28} = {\OpL[\text{SRR}, j]{4}} &
  = {-} (\bar d_j^\alpha \sigma_{\mu\nu} P_R \, d_i^\alpha) \,
    (\bar d_j^\beta  \sigma^{\mu\nu} P_R \, d_j^\beta) ,
\end{align}
{together with
\begin{align}
\OpL[\text{SRR}, j]{1} &
= (\bar d_j^\alpha P_R \, d_i^\beta) \, (\bar d_j^\beta P_R \, d_j^\alpha) ,
&
\OpL[\text{SRR}, j]{3} &
= {-} (\bar d_j^\alpha \sigma_{\mu\nu} P_R \, d_i^\beta) \,
  (\bar d_j^\beta  \sigma^{\mu\nu} P_R \, d_j^\alpha) .
\end{align}
Note that we choose the operators $Q_{26}$ and $Q_{28}$ with an opposite sign, compared to the basis in \cite{Aebischer:2021raf}.
For the SRR sector with four different quarks we define the structures}
\begin{equation}
  \label{eq:def-BMU-SRR-ops}
\begin{aligned}
  \OpL[\text{SRR}, Q]{1} &
  = (\bar d_j^\alpha P_R \, d_i^\beta) \, (\bar Q^\beta P_R \, Q^\alpha) , \qquad
&
  \OpL[\text{SRR}, Q]{3} &
  = {-} (\bar d_j^\alpha \sigma_{\mu\nu} P_R \, d_i^\beta) \,
    (\bar Q^\beta    \sigma^{\mu\nu} P_R \, Q^\alpha) ,
\\
  \OpL[\text{SRR}, Q]{2} &
  = (\bar d_j^\alpha P_R \, d_i^\alpha) \, (\bar Q^\beta P_R \, Q^\beta) ,
&
  \OpL[\text{SRR}, Q]{4} &
  = {-} (\bar d_j^\alpha \sigma_{\mu\nu} P_R \, d_i^\alpha) \,
    (\bar Q^\beta    \sigma^{\mu\nu} P_R \, Q^\beta) \,,
\end{aligned}
\end{equation}
{where the tensor structures have again opposite sign compared to the convention adopted in \cite{Aebischer:2021raf}. With these definitions we define the operators $Q_{29}-Q_{40}$}

\begin{equation}
\begin{aligned}
  (\OpL{29},\, \OpL{30},\, \OpL{31},\, \OpL{32}) &
  = (\OpL[\text{SRR}, u]{1},\, \OpL[\text{SRR}, u]{2},\,
     \OpL[\text{SRR}, u]{3},\, \OpL[\text{SRR}, u]{4}) ,
\\
  (\OpL{33},\, \OpL{34},\, \OpL{35},\, \OpL{36}) &
  = (\OpL[\text{SRR}, c]{1},\,\, \OpL[\text{SRR}, c]{2},\,\,
     \OpL[\text{SRR}, c]{3},\,\, \OpL[\text{SRR}, c]{4}) ,
\\
  (\OpL{37},\, \OpL{38},\, \OpL{39},\, \OpL{40}) &
  = (\OpL[\text{SRR}, d_k]{1},\, \OpL[\text{SRR}, d_k]{2},\,
     \OpL[\text{SRR}, d_k]{3},\, \OpL[\text{SRR}, d_k]{4}) .
\end{aligned}
\end{equation}

Finally, as far as the chirality-flipped operators are concerned, their
numbering in the BMU basis is given by
\begin{align}
  \OpL{40 + i} & = \OpL{i}[P_L \leftrightarrow P_R] ,
\end{align}
i.e. they are found by interchanging $P_L \leftrightarrow P_R$ in the
``non-flipped'' operators.

%
%
%

\section{Fierz identities} \label{subsec:fierz}
{In the process of transforming from} one operator basis to another one requires
the Fierz identities \cite{Fierz:1939zz} that
allow to transfer a given chain of spinors into  another
one. We list here the usual Fierz identities valid in $D=4$ dimensions that
we used in our analysis.

All  Fierz identities used are of the type $(12)(34) \to (14)(32)$
in which the exchange of fermion fields
$2\leftrightarrow 4$ (or equivalently $1\leftrightarrow 3$)
takes place. In the formulae below $P_A$ and $P_B$
stand for the usual projectors $P_{L,R}$  but in a given relation $P_A\not=P_B$. This means that if $P_A=P_L$, then $P_B=P_R$ and vice versa.

We have then
\be
(\bar\psi_1 P_A\psi_2)(\bar\psi_3 P_A\psi_4)=-\frac{1}{2}(\bar\psi_1 P_A\psi_4)(\bar\psi_3 P_A\psi_2)-\frac{1}{8}(\bar\psi_1 \sigma_{\mu\nu} P_A\psi_4)(\bar\psi_3 \sigma^{\mu\nu}P_A\psi_2),
\label{eq:fierzSRR}
\ee
\be
(\bar\psi_1 P_A\psi_2)(\bar\psi_3 P_B \psi_4)=-\frac{1}{2}(\bar\psi_1 \gamma_\mu P_B\psi_4)(\bar\psi_3 \gamma^\mu P_A\psi_2),
\ee
\be
\label{eq:fierzvll}
(\bar\psi_1\gamma_\mu P_A\psi_2)(\bar\psi_3 \gamma^\mu P_A \psi_4)=(\bar\psi_1\gamma_\mu P_A\psi_4)(\bar\psi_3 \gamma^\mu P_A \psi_2),
\ee
\be\label{eq:fierzVLR}
(\bar\psi_1\gamma_\mu P_A\psi_2)(\bar\psi_3 \gamma^\mu P_B \psi_4)=-2(\bar\psi_1 P_B\psi_4)(\bar\psi_3 P_A \psi_2),
\ee
\be
(\bar\psi_1 \sigma_{\mu\nu} P_A\psi_2)(\bar\psi_3 \sigma^{\mu\nu}P_A\psi_4)=-6 (\bar\psi_1 P_A\psi_4)(\bar\psi_3 P_A\psi_2)+\frac{1}{2}(\bar\psi_1 \sigma_{\mu\nu} P_A\psi_4)(\bar\psi_3 \sigma^{\mu\nu}P_A\psi_2),
\label{eq:fierzTRR}
\ee
\be
(\bar\psi_1 \sigma_{\mu\nu} P_A\psi_2)(\bar\psi_3 \sigma^{\mu\nu}P_B\psi_4)= 0\,.
\ee

For more Fierz identities involving charge conjugated fields see Appendix A.3 in
\cite{Buras:2020xsm}.

Apart from this we also need the color identity
\begin{equation}
\label{eq:colorid}
T^A \otimes T^A= \frac{1}{2} (\mathbb{\tilde 1} - \frac{1}{N_c} \mathbb{1}).
\end{equation}
%
\section{Master Formulae for One-loop Operator Insertions}
\label{app:master-formulae}
{In this section, we present master formulae for the 1-loop operator insertions. 
These can be used to obtain the shifts given in Sec.\ref{subsec:brules}.}
Consider a 4-fermion operator
\begin{equation} \label{eq:opgen}
	(\bar q_1 \hat V_1 \Gamma_1 q_2) (\bar q_3 \hat V_2 \Gamma_2 q_4)\,,
\end{equation}
here $\hat V_{1,2}$ and $\Gamma_{1,2}$ represent the color and Dirac structures.
There are two types of penguin insertions, an \emph{open penguin}, and the \emph{closed
penguin}. In the next two subsections we evaluate the corresponding amplitudes.

\subsection{Open penguin insertion}
The open penguin insertion of the operator \eqref{eq:opgen} gives
\begin{equation}
	P_{op} = W_{\lambda} ~ (-ig_s T^b \gamma_{\lambda^\prime}) ~
	\left ( \frac{-i g^{\lambda {\lambda^\prime} } }{q^2} \right )\,,
\end{equation}
\begin{equation}
	W_\lambda = i^2 \hat V_1 (-i g_s T^a) \hat V_2 I^{\mu \nu} T_{\mu \nu}^\lambda\,,
\end{equation}
where $T_{\mu \nu}^\lambda = \Gamma_1 \gamma_\nu \gamma_\lambda \gamma_\mu \Gamma_2$ and
\begin{equation} \label{eq:Imunu}
I^{\mu \nu}  = \int \frac{d^D k}{(2\pi)^D} \frac{k^\nu (k-q)^\mu}{k^2 (k-q)^2}
= -\frac{i}{16\pi^2} \frac{1}{\varepsilon} \left ( \frac{1}{6} q_\mu q_\nu + \frac{1}{12} q^2 g_{\mu \nu} \right )
-  \frac{i}{16\pi^2}  \left ( \frac{5}{18} q_\mu q_\nu + \frac{2}{9} q^2 g_{\mu \nu} \right ).
\end{equation}
Therefore the finite and infinite parts of $P_{op}$ can be written as
\begin{eqnarray}
	P_{op}(\rm infinite) &=& - C_{op} \frac{\alpha_s}{4\pi}\frac{1}{\varepsilon} \left ( \frac{1}{6} \frac{q_\mu q_\nu}{q^2}
	+\frac{1}{12} g_{\mu \nu}  \right ) \Gamma_1 \gamma_\mu \gamma_\lambda \gamma_\nu \Gamma_2 \otimes \gamma^\lambda\,,\\
	P_{op}(\rm finite) &=& - C_{op} \frac{\alpha_s}{4\pi} \left ( \frac{5}{18} \frac{q_\mu q_\nu}{q^2}
	+\frac{2}{9} g_{\mu \nu}  \right ) \Gamma_1 \gamma_\mu \gamma_\lambda \gamma_\nu \Gamma_2 \otimes \gamma^\lambda.
\end{eqnarray}
Here $C_{op}=\hat V_1 T^b \hat V_2 \otimes T^b$.

\subsection{Closed penguin insertion}
The closed penguin insertion of the operator \eqref{eq:opgen} gives
\begin{equation}
        P_{cl} = W_{\lambda} ~ (-ig_s T^b \gamma_{\lambda^\prime}) ~
        \left ( \frac{-i g^{\lambda {\lambda^\prime} } }{q^2} \right )\,,
\end{equation}
\begin{equation}
	W_\lambda =(-1) (i^2)(-i g_s)~ {\rm Tr} (\hat V_1 T^b ) \hat V_2
	~{\rm Tr}(\Gamma_1 \gamma_\mu \gamma_\lambda \gamma_\nu) \Gamma_2~ I^{\mu \nu}.
\end{equation}
Here the $I^{\mu \nu}$ is given by \eqref{eq:Imunu}.

Therefore the finite and infinite parts of $P_{cl}$ can be written as
\begin{eqnarray}
        P_{cl}(\rm infinite) &=& C_{cl} \frac{\alpha_s}{4\pi}\frac{1}{\varepsilon} \left ( \frac{1}{6} \frac{q_\mu q_\nu}{q^2}
	+\frac{1}{12} g_{\mu \nu}  \right ) {\rm Tr} (\Gamma_1 \gamma_\mu \gamma_\lambda \gamma_\nu) \Gamma_2
	\otimes \gamma^\lambda\,,\\
        P_{cl}(\rm finite) &=& C_{cl} \frac{\alpha_s}{4\pi}  \left ( \frac{5}{18} \frac{q_\mu q_\nu}{q^2}
        +\frac{2}{9} g_{\mu \nu}  \right ) {\rm Tr} (\Gamma_1 \gamma_\mu \gamma_\lambda \gamma_\nu) \Gamma_2
        \otimes \gamma^\lambda.
\end{eqnarray}
Here $C_{cl} = {\rm Tr} (\hat V_1 T^b) \hat V_2 \otimes T^b$.

\subsection{Special Cases}
In Tables~\ref{tab:tab2} and \ref{tab:tab3} we give the finite and 
singular parts for the penguin operator 
insertions with various Dirac structures.
\begin{table}[H]
\centering
\renewcommand{\arraystretch}{1.5}
\begin{tabular}{|l|l|l|}
\hline
Dirac Structure  & $P_{op}$(infinite)  &  $P_{op}$(finite) \\ \hline \hline
$\Gamma_1= \gamma_\rho P_L$, $\Gamma_2=\gamma^\rho P_L$ & $C_{op} \frac{\alpha_s}{4\pi} \gamma_\lambda P_L \otimes \gamma^\lambda$  &   -$C_{op} \frac{\alpha_s}{4\pi} \frac{13}{9} \gamma_\lambda P_L \otimes \gamma^\lambda$ \\ \hline
$\Gamma_1= \gamma_\rho P_L$, $\Gamma_2=\gamma^\rho P_R$ & 0  & 0  \\ \hline
$\Gamma_1=P_L$, $\Gamma_2=P_L$	& 0 &  0\\  \hline
$\Gamma_1=P_L$, $\Gamma_2=P_R$& $-C_{op} \frac{\alpha_s}{4\pi} \frac{1}{6}  \gamma_\lambda P_R\otimes \gamma^\lambda$  &
	$C_{op} \frac{\alpha_s}{4\pi} \frac{13}{8}  \gamma_\lambda P_R\otimes \gamma^\lambda$ \\ \hline
	$\Gamma_1=\sigma_{\alpha \beta } P_L$, $\Gamma_2= \sigma^{\alpha \beta} P_L$	& 0 & 0 \\ \hline
\end{tabular}
\caption{Finite and infinite parts of the open penguin insertion. $C_{op}=\hat V_1 T^b \hat V_2 \otimes T^b$.}
\label{tab:tab2}
\end{table}

\begin{table}[H]
\centering
\renewcommand{\arraystretch}{1.5}
\begin{tabular}{|l|l|l|}
\hline
Dirac Structure  & $P_{cl}$(finite)  &  $P_{cl}$(infinite) \\ \hline \hline
$\Gamma_1= \gamma_\rho P_L$, $\Gamma_2=\gamma^\rho P_L$ & - $C_{cl} \frac{\alpha_s}{4\pi} \frac{13}{9} \gamma_\lambda P_L \otimes \gamma^\lambda$  &   $C_{cl} \frac{\alpha_s}{4\pi} \frac{1}{3} \gamma_\lambda P_L \otimes \gamma^\lambda$ \\ \hline
	$\Gamma_1= \gamma_\rho P_L$, $\Gamma_2=\gamma^\rho P_R$ & -$C_{cl} \frac{\alpha_s}{4\pi} \frac{13}{9} \gamma_\lambda P_R \otimes
	\gamma^\lambda$   & $C_{cl} \frac{\alpha_s}{4\pi} \frac{1}{3} \gamma_\lambda P_R \otimes
        \gamma^\lambda$   \\ \hline
$\Gamma_1=P_L$, $\Gamma_2=P_L$  & 0 &  0\\  \hline
$\Gamma_1=P_L$, $\Gamma_2=P_R$& 0  & 0 \\ \hline
$\Gamma_1=\sigma_{\alpha \beta } P_L$, $\Gamma_2= \sigma^{\alpha \beta} P_L$    & 0 & 0 \\ \hline
\end{tabular}
\caption{Finite and infinite parts the of closed penguin insertion.
$C_{cl} = {\rm Tr} (\hat V_1 T^b) \hat V_2 \otimes T^b$.}
\label{tab:tab3}
\end{table}

%
%

\renewcommand{\refname}{R\lowercase{eferences}}

\addcontentsline{toc}{section}{References}

\bibliographystyle{JHEP}

\small

\bibliography{Bookallrefs}

\end{document}